\documentclass[preprint,onecolumn,amsmath,letterpaper,amssymb,prl,floatfix,superscriptaddress]{revtex4-1}
\usepackage{graphicx}
\usepackage{color}
\usepackage{chemmacros}
\usepackage{chemarr}
\usepackage{filecontents}
\usepackage[titletoc,toc,title]{appendix}
\begin{document}
\bibliographystyle{nature1}

\title{Optimal information transfer in enzymatic networks: A field theoretic formulation}

\author{Himadri S. Samanta}\affiliation{Department of Chemistry, The University of Texas at Austin, TX 78712}
%\affiliation{Biophysics Program, Institute For Physical Science and Technology, University of Maryland, College Park, Md 20742}
\author{Michael Hinczewski} 
\affiliation{Department of Physics, Case Western Reserve University, OH 44106}
\author{D. Thirumalai} 
\affiliation{Department of Chemistry, The University of Texas at Austin, TX 78712}
%\affiliation{Biophysics Program, Institute For Physical Science and Technology, University of Maryland, College Park, Md 20742}
\date{\today}
\begin{abstract}
Signaling in enzymatic networks is typically triggered by
environmental fluctuations, resulting in a series of stochastic
chemical reactions, leading to corruption of the signal by noise. For
example, information flow is initiated by binding of extracellular
ligands to receptors, which is transmitted through a
{cascade involving} kinase-phosphatase stochastic chemical
reactions. For a class of such networks, we develop a general
field-theoretic approach in order to calculate the error in signal
transmission as a function of an appropriate control variable.
Application of the theory to a simple push-pull network, a module in
the kinase-phosphatase cascade, recovers the exact results for error
in signal transmission previously obtained using umbral calculus
(Phys. Rev. X., {\bf 4}, 041017 (2014)). We illustrate the generality
of the theory by studying the minimal errors in noise reduction in a
reaction cascade with two connected push-pull modules. Such a cascade
behaves as an effective {three-species} network with a pseudo
intermediate. In this case, optimal information transfer, resulting in
the smallest square of the error between the input and output, occurs
with a time delay, which is given by the inverse of the decay rate of
the pseudo intermediate. Surprisingly, in these examples the minimum
error computed using simulations that take non-linearities and
discrete nature of molecules into account coincides with the
predictions of a linear theory. In contrast, there are substantial
deviations between simulations and predictions of the linear theory in
error in signal propagation in an enzymatic push-pull network for a
certain range of parameters. Inclusion of second order perturbative
corrections shows that differences between simulations and theoretical
predictions are minimized. Our study establishes that {a} field
theoretic formulation {of} stochastic biological {signaling
offers} a systematic way to understand error propagation in networks of arbitrary complexity.
\end{abstract}

\maketitle
\def\s{\rule{0in}{0.28in}}

\section{Introduction:}

Cell signaling involves the ability of cells to detect changes in the environment and respond to them~\cite{Goldbeter81PNAS, Thattai01PNAS,Thattai02BJ,Eldar:2010aa, Arjun08Cell,Maheshri:2007aa}, a fundamental necessity of living systems. Several signaling networks involve proteins, which switch between active and inactive states. By quantitatively describing how different signaling proteins are functionally linked, we can understand the behavior of signaling pathways, and the associated bandwidth that determines fidelity of information transfer~\cite{Bowsher:2013aa}. 
%Phosphorylation to activate kinases and dephosphorylation to
%inactivate them is a central mechanism in cell
%signaling~\cite{Levine:aa,Stadtman:1977aa,Goldbeter81PNAS,Detwiler:aa,Heinrich:aa}. Alterations
%of a proteins function are often, although not always, brought about
%by the addition or removal of a phosphate group: phosphorylation and
%dephosphorylation. As with most signaling mechanisms, dysfunction of
%phosphorylation can lead to many diseases such as
%cancer~\cite{Kolch:2015aa}. Consequently, a vast number of drugs are
%directed against the modulation of phosphorylation of proteins in
%cells.
In typical enzymatic networks, environmental information is
transmitted into the cell interior through cascades of stochastic
biochemical reactions~\cite{cai08Nature}. Noise inevitably propagates
through the cascade, potentially corrupting the signal.  Depending on
the parameters, small changes in the input can be translated into
large (but noise corrupted) output variations. The amplification is
essential but it must also preserve the signal content to be useful
for downstream processes. The signaling circuit, despite operating in
a noisy environment, needs to maintain high fidelity between output
and the amplified input~\cite{Becker15PRL}. Over the years concepts in
information theory have been adopted to assess the fidelity of signal
transmission in the context of biochemical
network~\cite{Lestas10Nature,Bode50PIRE}. Several studies have used
mutual information between input and output signals to quantify the
reliability of signal
transduction~\cite{deRonde10PRE,Ziv07PlosOne,Tkacik08PRE,Walczak09PNAS,Mehta09MolCellBiol,Mugler09PRE}. The
formalism has been applied to the study of a variety of networks
including cascades and networks with
feedback~\cite{deRonde10PRE}. These and other studies have expanded
over understanding of the fidelity of information transfer in
biological networks in which both noise and copy number fluctuations
are important.

In a recent paper~\cite{MH14PRX}, we considered the problem of how to
extract information faithfully from noisy signals using mathematical
methods developed in the context of communication theory developed
over sixty years ago by Wiener~\cite{Wiener49} and independently by
Kolmogorov~\cite{Kolmogorov41}. {The Wiener-Kolmogorov (WK)
  approach has since proven a useful tool in a variety of contexts in
  biological signaling~\cite{MH16JPCB,Becker15PRL,Hathcock16}.  The WK
  theory,} reformulated by Bode and Shannon~\cite{Bode50PIRE}, assumes
that the input and output are continuous variables that describe
stationary stochastic processes. {The goal of approach is to
  minimize the mean squared error between the input and output
  signals, but the optimization is restricted to the space of only
  linear noise filters.} Recently, we developed an analytic formalism
of general validity to overcome {some of the} limitations of the WK theory based
on exact techniques involving umbral calculus~\cite{Roman05}. We
illustrated the efficacy of the non-linear theory with applications to
push-pull network and its variants including instances when the input
is time-dependent.

The use of non-standard mathematics in the form of umbral calculus,
perhaps, obscures the physics of optimal filtering in biological
networks in which the effects of non-linearities in signal
amplification have to be considered. Here, we develop an alternate
general formalism based on a many body formulation of reaction
diffusion equations introduced by Doi and Peliti~\cite{doi76JPA,
  peliti85JP}. This formulation converts the signal optimization
problem to a standard field theory, allowing us to calculate the
response and correlation functions by standard methods. The advantage
of this formalism is that both discrete and continuum cases can be
studied easily. Non-linear contributions can be obtained using
systematic diagrammatic perturbation scheme for an arbitrary
network. {Networks where} temporal dynamics {are} coupled with
spatial gradients in signaling activities, which {regulate}
intracellular processes and signal propagation across the cell, can
also be investigated using the present formalism. Application of the
theory to a push-pull network and a simplified biochemical network
recovers the exact results obtained in our previous study. We also
extend the formalism to solve signal transduction in a cascade, which
serves as a {model} for a variety of biological networks.  The
formalism is general and is applicable to arbitrary networks with
feedback, time delay and special variations~\cite{Silva:aa}. {Our
  work exploits} standard methods in physics, {illustrating the
  usefulness of a field theoretic formulation at the interface of
  communication theory and biology.}

 %We showed how a simple kinase-phosphatase push-pull network, a basic unit of signaling pathway behaves as a Wiener-Kolmogorov (WK) optimal noise filter\cite{2,3}.
%Optimality has been realized by tuning phosphatase levels, which we verified using simulations of a microscopic model of the loop reaction network. The optimality is robust, with the filter operating at near-optimal levels even when the WK conditions are only approximately fulfilled, over a broad range of realistic parameter values. 

%We study the effective noise reduction in signaling pathways. We see it behaves like WK optimal filter. We get bounds and conditions for optimal signal transduction with the help of Wiener-Kolmogorov filter theory. Optimality has been realized in a stochastic simulation of the full loop reaction system with realistic parameters. We calculate relative error between time varying output and input signal using field theory for stochastic interacting particle systems\cite{4}. Minimum of relative error has been computed, which is exactly matches with the WK minimum.

\section{Theory:}

\paragraph{\bf Linear Push-Pull Network:}
In order to develop the many body formalism for {a} general signaling
network, we first consider a simple model. The concepts and the
general diagrammatic expansion developed in this context, lays the
foundation for applications to more complicated enzymatic networks as
well as signaling cascades.  In a typical signaling pathway, for
example the {mitogen activated protein kinase
(MAPK)~\cite{Schoeberl02Nature,Hersen08PNAS} pathway}, external and
environmental fluctuations activate a cascade of enzymatic reactions,
thus transmitting information across the membrane in a sequential
manner. Each step involves activation of kinases by phosphorylation
reaction and deactivation by
phosphatases~\cite{Levine:aa,Stadtman:1977aa,Detwiler:aa,Heinrich:aa,Kolch:2015aa}.
A truncated version of such a cascade is a single step (Fig.1), which
we refer to as a push-pull network~\cite{MH14PRX}. In this signaling
network, there are only two chemical species. One is $I(t)$ (the
"input") and the other is $O(t)$ (the "output") whose production
depends on $I(t)$.  The upstream pathway, which serves as an external
signal, creates the species $I$ by the reaction $\phi
\xrightarrow[]{F} I$ with an effective production rate $F$. The output
$O$ is a result of the reaction $I \xrightarrow[]{R(I)} I+O$, with a
rate $R(I(t))$ that depends on the input. The species are deactivated
through $I\xrightarrow[]{\gamma_I}\phi$ and $O
\xrightarrow[]{\gamma_O}\phi$ with rates $\gamma_I$ and $\gamma_O$
respectively, mimicking the role of phosphatases (Fig.1). The input
varies over a characteristic time scale $\gamma_I^{-1}$, fluctuating
around the mean value $\bar{I}=F/ \gamma_I$. The degradation rate sets
the time scale $\gamma_O^{-1}$ over which $O(t) $ responds to changes
in the input.

The chemical Langevin equations describing the changes in $I$ and $O$ are,
\begin{equation}\label{langevin}
\frac{dI}{dt}=F-\gamma_I I + \eta_I, \ \ \frac{dO}{dt}=R(I)-\gamma_O O + \eta_O,
\end{equation}
where $\eta_I$ and $\eta_O$ are Gaussian white noise with zero mean
($\langle\eta_\alpha\rangle=0$) and correlation $\textless \eta_\alpha(t)
\eta_\alpha' (t')\textgreater=2 \sqrt{\gamma_\alpha
  \bar{\alpha}}\delta_{\alpha \alpha'}\delta(t-t')$ with $\alpha=I,O$
and $\bar{\alpha}$ is the mean population $\alpha$. For small
fluctuations, $\delta \alpha(t)=\alpha(t)-\bar{\alpha}$,
Eq.(\ref{langevin}) can be solved using a linear approximation for the
rate function $R(I(t))\approx R_0 \bar{I} + R_1 \delta I(t)$, with
coefficients $R_0, R_1 \textgreater 0$. The result is
\begin{eqnarray}\label{sol}
\delta I(t) &=& \int_{-\infty}^t dt' e^{-\gamma_I (t-t')} \eta_I (t'), \\ \nonumber 
\delta O(t) &=& \int_{-\infty}^t dt' \frac{R_1}{G} e^{-\gamma_O (t-t')}\left [ G \delta I(t') +\frac{G}{R_1} \eta_O (t')\right ],
\end{eqnarray}
where in the second line an arbitrary scaling factor $G$ has been
introduced. The solution for $\delta O(t)$ has the structure of a
linear noise filter equation; $\tilde{s}=\int_{-\infty}^t dt'
H(t-t')c(t')$, with $c(t)=s(t)+n(t)$. The signal $s(t)=G\delta I(t)$
together with the noise term $ n(t)\equiv G R_1^{-1} \eta_O (t)$
constitute the corrupted signal, $c(t)$. The output $\tilde{s}
(t)\equiv \delta O(t) $ is produced by convolving $c(t)$ with a linear
kernel $H(t) \equiv R_1 G^{-1} \exp(-\gamma_O t)$, which filters the
noise. As a consequence of causality, the filtered output $\tilde{s}$
at time $t$ depends only on $c(t')$ from the past.

The primary goal in transmitting signal with high fidelity is to
devise an optimal causal filter, $H_{opt}(t)$, which renders
$\tilde{s} (t)$ as close to $s(t)$ as possible. In a remarkable
development, Weiner~\cite{Wiener49} and Kolmogorov~\cite{Kolmogorov41}
independently discovered a solution to this problem {in} the context
of communication theory, which launched the modern era in signal
decoding from time series. In particular, WK proposed a solution that
minimizes the square of the differences between $\tilde{s}$ and $s(t)$
by seeking an optimal filter $H_{WK}(t)$ among all possible linear
filters. In the push-pull network, this means having $\delta O(t)$
reproduce as accurately as possible the scaled input signal $ G \delta
I(t)$. For a particular $\delta I(t)$ and $\delta O(t)$, the value of
the mean squared error $E= \langle(\tilde{s} -s)^2\rangle/\langle s^2\rangle$ is smallest when
$G=\langle(\delta O)^2\rangle/\langle\delta O \delta I\rangle$, which we identify as a gain
factor. In this case, $E= 1-\langle\delta O \delta I\rangle^2/(\langle(\delta
O)^2\rangle\langle(\delta I)^2\rangle)$.

The optimal causal filter $H_{WK}$ satisfies the following Wiener-Hopf
equation\cite{MH14PRX,Becker15PRL},
\begin{equation}\label{WK}
C_{cs}(t)=\int_{-\infty}^t dt' H_{WK}(t-t')C_{cc}(t'), \ \ t\rangle0
\end{equation}
where $C_{xy}(t)\equiv \langle x(t')y(t'+t)\rangle$ is the correlation
between points in the time series $x$ and $y$, assumed to depend only
on the time difference $t-t'$. We can evaluate the correlation
functions $C_{cs}$ and $C_{cc}$ using Eq.~(\ref{sol}), and
substituting these solutions in Eq. (\ref{WK}), the optimal filter
function can be solved by assuming a generic ansatz,
$H_{WK}(t)=\sum_{i=1}^N A_i \exp(-\lambda_i t)$. The unknown
coefficients, $A_i $, and the associated rate constants $\lambda_i$
are found by comparing the left and right hand sides of
Eq. (\ref{WK}). Elsewhere~\cite{MH14PRX}, we showed that $H_{WK }(t)=
\gamma_I (\sqrt{1+\Lambda}-1) \exp(\gamma_I \sqrt{1+\Lambda}t)$.  The
conditions for achieving WK optimality, $H(t)=H_{WK}(t)$,
are~\cite{MH14PRX},
\begin{equation}
\gamma_O=\gamma_I \sqrt{1+\Lambda}, \ \ G=\frac{R_1}{\gamma_I(\sqrt{1+\Lambda}-1)},
\end{equation}
leading to the minimum relative error,
\begin{equation}\label{optimal}
E_{WK}=\frac{2}{1+\sqrt{1+\Lambda}}, ~~\Lambda \equiv \frac{R_1^2}{R_0 \gamma_I}.
\end{equation} 
The fidelity between the output and input is described through a single dimensionless optimality control parameter, $\Lambda$, which can be written as 
$\Lambda \equiv (R_0/\gamma_I)(R_1/R_o)^2$. The first term, $R_0/\gamma_I$, is a burst factor, measuring the mean number of output molecules produced per input molecule during the active lifetime of the input molecule. The second term, $(R_1/R_o)^2$, is a sensitivity factor, reflecting the local response of the production function $R(I)$ near $\bar{I}$ (controlled by the slope $R_1 = R^\prime(\bar{I})$) relative to the production rate per input molecule $R_0=R(\bar{I})/\bar{I}$. 

In our recent work~\cite{MH14PRX}, we extended the WK approach to
include non-linearity and {the} discrete nature of the input and output
molecules $I$ and $O$ \cite{MH14PRX}. Both these considerations are
relevant in biological circuits where $R(I)$ is non-linear and the
copy numbers of $I$ and $O$ are likely to be small. Starting from the
exact master equation, valid for discrete populations and arbitrary
$R(I)$, we rigorously solved the original optimization problem for the
error $E$ between output and input using the principles of umbral
calculus\cite{Roman05}. The main results are as follows.  For any
arbitrary function expanded as,
\begin{equation}\label{seriesUC}
R(I)= \sum_{n=0}^\infty \sigma_n v_n(I),
\end{equation} 
with $v_n(I)=\sum_{m=0}^{\infty} (n-m)! (-\bar{I})^m \begin{pmatrix} n \\ m \end{pmatrix} \begin{pmatrix} I \\ n-m \end{pmatrix} $ and $\sigma_n=\textless v_n(I) R(I)\textgreater/(\bar{I}^n n!)$, the relative error can be expressed by an exact expression, 
\begin{equation}\label{exact}
E=1-\frac{\bar{I} \gamma_O^2 \sigma_1^2}{(\gamma_I+\gamma_O)^2}\left[ \gamma_O \sigma_0 +\sum_{n=1}^{\infty} \sigma_n^2 \frac{n!\gamma_O \bar{I}^n}{\gamma_O+n\gamma_I}\right]^{-1}.
\end{equation}
The expression above is bounded from below by
  \begin{equation}
 E \ge E_{opt} \equiv \frac{2}{1+\sqrt{1+\tilde{\Lambda}}},
 \label{EKMIN}
 \end{equation} 
where $\tilde{\Lambda}=\bar{I} \sigma_1^2/(\sigma_0 \gamma_I)$. The
equality is only reached when $\gamma_O=\gamma_I
\sqrt{1+\tilde{\Lambda}}$ and $R(I)$ is an optimal linear filter of
the form, $R_{opt}(I)=\sigma_0 + \sigma_1 (I-\bar{I})$, with all
$\sigma_n=0$ for $n\ge 2$. Obtaining the lower bound is important for
noise reduction in biological networks as it provides insights into
energy costs required to reduce the error~\cite{Lestas10Nature}.
 
 \paragraph{\bf Field theoretic formulation:}
 In order to generalize the results in our previous
 study~\cite{MH14PRX} to arbitrary regulatory networks, we adopt a
 many body approach pioneered by Doi and Peliti\cite{doi76JPA,
   peliti85JP}. Such an approach has been used in the study of a
 variety of reaction diffusion equations~\cite{Lee95JSP,
   Cardy98JSP}. Besides suggesting plausible new ways of examining how
 {signals are} transmitted in biochemical reaction networks, the current
 theory shows how standard field theoretic methods can be adopted for
 use in control theory. By way of demonstrating its utility, we {rederive}
 the exact analytical solution (Eq.(\ref{exact})) for the relative
 error in the push-pull network.~In the Doi-Peliti formalism, the
 configurations at time $t$ in a locally interacting many body
 system {are} specified by the occupation numbers of each species on a
 lattice site $i$. In our case, $I_i $ is the input population and
 $O_i $ is the output population. As a consequences of the stochastic
 dynamics, the on-site occupation numbers are modified. Arbitrarily
 many particles of either {population} are allowed to occupy any
 lattice site. In other words, $I_i$, $O_i=0,1,\cdots \infty$. The
 master equation for the local reaction scheme that governs the time
 evolution of the configurational probability with $I_i$ input and
 $O_i$ output at site $i$ at time $t$ is obtained through the balance
 of gain and loss terms. The result is,
\begin{eqnarray}\label{a}
\frac{dP(I_i,O_i,t)}{dt}&=&\gamma_I [(I_i+1) P(I_i+1,O_i,t)-I_i P(I_i,O_i,t)]+F[P(I_i-1,O_i,t)-P(I_i,O_i,t)] \\ \nonumber
&+&\gamma_O[(O_i+1)P(I_i,O_i+1,t)-O_iP(I_i,O_i,t)]+R(I_i)[P(I_i,O_i-1,t)-P(I_i,O_i,t)]. \\ \nonumber
\end{eqnarray}
We use {the} Fock space representation to account for the changes in the
site occupation number by integer values for the chemical reactions
describing the network. Following Doi and Peliti, we introduce the
{bosonic} ladder operator algebra with commutation relation
$[a_i,a_j]=0$, $[a_i,a_j^\dagger]=\delta_{ij}$ for the input
population, allowing us to construct the input particle number
eigenstates $| I_i \rangle$ obeying $a_i | I_i \rangle=I_i |
I_i -1 \rangle$, $a_i^\dagger | I_i \rangle=| I_i
+1\rangle$, $a_i^\dagger a_i | I_i \rangle=I_i |I_i
\rangle$. A Fock state with $I_i $ particles on site $i$ is
obtained from the vacuum state $|0\rangle$, defined by the
relation $a_i |0\rangle=0$, and $|I_i
\rangle={a_i^\dagger}^{I_i} |0\rangle$. Similarly, we
introduce annihilation and creation operators for output particles
$b_i$ and $b_i^\dagger$ that commute with the input ladder operators:
$[a_i, b_j]=0=[a_i,b_j^\dagger]$.

Stochastic kinetics for the entire lattice is implemented by
considering the master equation for the configurational probability
$P(\{I_i\},\{O_j\},t)$, given by a sum over all lattice points on the
right hand side of Eq.(\ref{a}), by noting that a general Fock state
is constructed by the tensor product $| \{I_i\},\{O_j\} \rangle=\Pi_i
|I_i \rangle |O_i \rangle$. We define a time dependent formal state
vector through a linear combination of all possible Fock states,
weighted by their configurational probability at time $t$,
\begin{equation}
| \psi(t)\rangle=\sum_{\{I_i,O_j\}}^{\infty}P(\{I_i\},\{O_j\},t)\Pi_{\{i,j\}}^{}{a^{\dagger}_i}^{I_i}{b^{\dagger}_i}^{O_j}|0\rangle.
\end{equation}
This superposition state encodes the stochastic temporal evolution. We
use standard methods to transform the time dependence from the linear
master equation into an imaginary time Schr\"{o}dinger equation,
governed by a time-dependent stochastic evolution operator $H$,
\begin{equation}\label{sch}
\frac{d}{dt} |
\psi(t)\rangle=-H(\{a\},\{a^{\dagger}\};\{b\},\{b^{\dagger}\}) |
\psi(t)\rangle.
\end{equation}
We may multiply Eq.(\ref{a}) by
${a_i^{\dagger}}^{I_i}{b_i^{\dagger}}^{O_i} | 0\rangle$, and sum over
all values of $I_i,O_i$.  With the definition of the state $|
\psi(t)\rangle$,
\begin{equation}
| \psi(t)\rangle=\sum_{I_i,O_i}^{\infty}P(I_i,O_i,t){a_i^{\dagger}}^{I_i}{b_i^{\dagger}}^{O_i}| 0\rangle,
\end{equation}
the $\gamma_I$ term, i.e. $\gamma_I [(I_i+1) P(I_i+1,O_i,t)-I_i P(I_i,O_i,t)]$,  in Eq.(\ref{a}) becomes,
\begin{eqnarray}
&&\gamma_I \sum_{I_i,O_i}^{\infty} [(I_i+1) P(I_i+1,O_i)-I_i P(I_i,O_i)]{a_i^{\dagger}}^{I_i}{b_i^{\dagger}}^{O_i} | 0\rangle \nonumber \\
&&=\gamma_I \sum_{I_i,O_i}^{\infty}[P(I_i+1,O_i)a_i {a_i^{\dagger}}^{I_i+1}{b_i^\dagger}^{O_i} | 0\rangle-P(I_i,O_i)a_i^\dagger a_i {a_i^{\dagger}}^{I_i}{b_i^\dagger}^{O_i} | 0\rangle ].
\end{eqnarray}
By relabeling the indices in the first sum, we arrive at the desired Hamiltonian expressed in second quantized representation as, 
%\begin{equation}
$H_{\gamma_I}=-\gamma_I (1-a_i^\dagger)a_i.$
%\end{equation} 
Similarly, terms with coefficients $F$, $\gamma_O$ and $R(I_i)$  in Eq.(\ref{a}) give the following contributions, 
%\begin{eqnarray}
$H_{F}=-F (a_i^\dagger-1),~
H_{\gamma_O}=-\gamma_O (1-b_i^\dagger)b_i,~  
H_{R(I_i)}=-R(a_i^\dagger a_i) (b_i^\dagger-1). $
%\end{eqnarray}
The total Hamiltonian $H$ takes the following form,
\begin{eqnarray}\label{HH}
H&=&H_{\gamma_I}+H_{F}+H_{\gamma_O}+H_{R(I)} \nonumber \\
&=&\sum_i [-\gamma_I (1-a_i^\dagger)a_i-F (a_i^\dagger-1)-\gamma_O (1-b_i^\dagger)b_i-R(a_i^\dagger a_i) (b_i^\dagger-1)].
\end{eqnarray}

A convenient choice for the initial configuration for the master equation describing the stochastic particle reactions is an independent Poisson distribution at each site,
\begin{equation}
P(\{I_i\},\{O_i\};0)=\Pi_i P_0(I_i) P_0(O_i)=\Pi_i e^{-\bar{I}_0}
e^{-\bar{O}_0}{O_0}^{-O_i} {I_0}^{-I_i}/I_i!O_i!.
\end{equation}
with mean initial input and output concentrations $\bar{I_0}$ and $\bar{O}_0$.
Just as in quantum mechanics, Eq.(\ref{sch}) can be formally solved leading to,
\begin{equation}
\mid \psi(t) \rangle=e^{H t} | \psi(0)\rangle,
\end{equation}
with the initial state $| \psi \rangle=e^{\bar{I}_0 \sum_i (a_i^\dagger-1)+\bar{O}_0 \sum_i (b_i^\dagger-1)}|0\rangle$.

Our goal is to compute averages and correlation functions with respect
to the configurational probability $P(\{I_i\},\{O_i\};t)$, which is
accomplished by means of the projection state
$\textless\mathcal{P}|=\textless0| \Pi_i e^{a_i +b_i}$, for which
$\textless \mathcal{P}|0\rangle=1$ and $\textless \mathcal{P}|
a_i^\dagger=\textless \mathcal{P}|=\textless \mathcal{P}|b_i^\dagger$,
since $[e^{a_i},a_j^\dagger]=e^{a_i}\delta_{ij}$.  The average value
of an observable $A(\{I_i\},\{O_i\})$ is,
\begin{equation}
\textless A(t)\rangle=\sum_{\{I_i\},\{O_i\}}A(\{I_i\},\{O_i\})P(\{I_i\},\{O_i\};t),
\end{equation}
from which the statistical average of an observable can be calculated using,
\begin{eqnarray}
\textless A(t)\rangle&=&\textless\mathcal{P}| A(\{a_i^{\dagger},a_i;b_i^\dagger,b_i\})| \psi(t)\rangle \\ \nonumber
&=&\textless \mathcal{P}| A(\{a_i^{\dagger},a_i;b_i^\dagger,b_i\})e^{-H((\{a_i^{\dagger}\},\{a_i\};\{b_i^\dagger\},\{b_i\})t}| \psi(0)\rangle.
\end{eqnarray}

We follow a well-established route in quantum many particle
theory~\cite{Negele88}, and proceed towards a field theory
representation by constructing a path integral equivalent of the time
dependent Schr\"{o}dinger equation (Eq.(\ref{sch})) based on coherent
states~\cite{Tauber14}. These are defined as right eigenstates of the
annihilation operators, $a_i |\alpha_i \rangle=\alpha_i |\alpha_i
\rangle$ and $a_i |\beta_i \rangle=\beta_i |\beta_i \rangle$, with
complex eigenvalues $\alpha_i$ and $\beta_i$. The coherent states
satisfy $|\alpha_i \rangle= \exp(\frac{1}{2}|\alpha_i|^2+\alpha_i
\alpha_i^\dagger)|0\rangle$, the overlap integral $\textless\alpha_j
|\alpha_i
\rangle=\exp(-\frac{1}{2}|\alpha_i|^2-\frac{1}{2}|\alpha_j|^2+\alpha_j^*
\alpha_i)$, and the completeness relation $\int \Pi_i d^2 \alpha_i
|\{\alpha_i\}\rangle \textless \{\alpha_i\}|=\pi$. After splitting the
temporal evolution (Eq.(\ref{sch})) into infinitesimal increments,
inserting the completeness relation at each time step, and with
additional manipulations leads to an expression for the
configurational average,
\begin{equation}
\langle A(t)\rangle \propto \int \Pi_i d\alpha_i d \alpha_i^* d
\beta_i d\beta_i^*
A(\{\alpha_i\},\{\beta_i\})e^{-\mathcal{S}[\alpha_i^*,\beta_i^*,\alpha_i,\beta_i]}.
\end{equation}
The exponential statistical weight is determined by the action,
\begin{equation}\label{action}
\mathcal{S}[[\alpha_i^*,\beta_i^*,\alpha_i,\beta_i]=\sum_i \left [
    \int_0^{t_f} \left\{ \alpha_i^* (t) \frac{\partial
      \alpha_i(t)}{\partial t}+\beta_i^* (t) \frac{\partial
      \beta_i(t)}{\partial t} \right \}
    +H(\alpha_i^*,\beta_i^*,\alpha,\beta) \right ] dt.
\end{equation}
Finally, by taking the continuum limit using $\sum_i \rightarrow
a_0^{-d} \int d^d x$, $a_0$ is a {lattice} constant, $\alpha_i
(t)\rightarrow \phi (x,t)$, $\beta_i (t)\rightarrow \psi (x,t)$ and
$\alpha_i (t)\rightarrow a_0^d \phi (x,t)$, $\beta_i^* (t)\rightarrow
a_0^d \psi^* (x,t)$, the expectation value is represented by a
functional integral,
\begin{equation}
\langle A(t)\rangle \propto \int \Pi_i \mathcal{D}[\phi^*,\phi,\psi^*,\psi] A(\{\phi\},\{\psi\})e^{-\mathcal{S}[\psi^*,\phi^*,\psi,\phi]}
\end{equation}
with an effective action
\begin{equation}\label{Action}
\mathcal{S}[\psi^*,\phi^*,\psi,\phi]=\int_0^{t_f}\left[\left\{ \psi^* (t) \frac{\partial \psi(t)}{\partial t}+\phi^* (t) \frac{\partial \phi(t)}{\partial t} \right \} +H(\psi^*,\phi^*,\psi,\phi) \right ] dt.
\end{equation}
 In the  Hamiltonian (Eq.(\ref{HH})), $a^\dagger$ and  $b^\dagger$ are
 replaced   by   the   field    variables   $\phi^*$   and   $\psi^*$,
 respectively.  Similarly, $a$  and  $b$ operators  become $\phi$  and
 $\psi$ respectively.

The action in Eq.(\ref{Action}) encodes the stochastic master equation
kinetics through four independent fields
($\psi^*,\phi^*,\psi,\phi$). With this formulation, an immediate
connection can be made to the response functional formulation using
the Janssen - De Dominicis formalism for Langevin equations
\cite{Dominicis76JPC,Janssen76ZPB}. In this approach, the response
field enters at most quadratically in the pseudo-Hamiltonian, which
may be interpreted as averaging over Gaussian white noise. With this
in mind, we apply {the} non-linear Cole-Hopf transformation \cite{Cole51QAM,
  Hopf50CPAM}, in order to obtain quadratic terms in auxiliary fields,
%\begin{eqnarray}
$\phi^*= e^{\bar{\phi}_I}, \  \phi=e^{-\bar{\phi}_I} \phi_I,  \
\psi^*= e^{\bar{\psi}_O},  \  \psi=e^{-\bar{\psi}_O} \psi_O$,
%\end{eqnarray}
to the action in Eq.(\ref{Action}). The Jacobian for this variable
transformation is unity, and the local particle densities are $\phi^*
\phi =\phi_I$ and $\psi^* \psi =\psi_O$. We obtain the following
Hamiltonian,
\begin{equation}
H=-\gamma_I (-\bar{\phi}_I+\frac{\bar{\phi}_I^2}{2} )\phi_I
-F(\bar{\phi_I}+\frac{\bar{\phi}_I^2}{2} )
-\gamma_O (-\bar{\psi}_O+\frac{\bar{\psi}_O^2}{2} )\psi_O
-R(\phi_I)(\bar{\psi}_O+\frac{\bar{\psi}_O^2}{2} ).
\end{equation}
In the above equation, {the} exponential term has been expanded to
second order. The rate equations are obtained through $\delta
\mathcal{S}/ \delta \bar{\psi} \mid_{\bar{\psi}=0}=0$ and $\delta
\mathcal{S}/ \delta \bar{\phi} \mid_{\bar{\phi}=0}=0$.  The terms
quadratic in the auxiliary fields~($\bar{\psi}$ and $\bar{\phi}$)
encapsulate the second moment of the Gaussian white noise with zero
mean.

In order to obtain fluctuation corrections needed to calculate minimum
error in signal transduction, we write the action in terms of
fluctuating fields, $\delta \phi_I =\phi_I-\langle\phi_I\rangle$ and
$\delta \psi_O =\psi_O-\langle\psi_O\rangle$ as,
\begin{eqnarray}
H&=&\bar{\phi}_I[\gamma_I \delta \phi_I- \gamma_I \langle\phi_I\rangle\bar{\phi}_I]+\nonumber \\ &&
\bar{\psi}_O [\gamma_O \delta \psi_O -\{c_1 \delta \phi_I +\frac{c_2}{2}\delta \phi_I^2 +\cdots\}-\gamma_O \langle\psi_O\rangle\bar{\psi}_O]
\end{eqnarray}
where we have expanded $R(\phi_I)$ in a Taylor series,
\begin{equation}\label{rphi}
 R(\phi_I)=\sum_0^\infty \frac{c_n}{n!}(\delta \phi_I)^n,
 \end{equation}
with constant $c_n$. Note this expansion differs from the one used in Eq.(\ref{seriesUC}). The coefficients of $\bar{\phi}_I^2$ and $\bar{\psi}_O^2$ reflect the noise correlations in Langevin description. %Here, we omit the terms ($\propto \delta \phi$ and higher order terms) in the Langevin noise contribution.

In Fourier space the action becomes
\begin{eqnarray}\label{action} 
\mathcal{S}[\tilde{\Psi}, \Psi]&=&\int_w \bar{\phi}_I[-iw \  \delta \phi_I+\gamma_I \delta \phi_I- \gamma_I \langle\phi_I\rangle\bar{\phi}_I]+\\ \nonumber
&&\bar{\psi}_O [-iw \ \delta \psi_O+\gamma_O \delta \psi_O -c_1 \delta \phi_I - \gamma_O \langle\psi_O\rangle\bar{\psi}_O]\\ \nonumber
&&+\mathcal{S}_{int}[\tilde{\Psi}, \Psi]
\end{eqnarray}
where $\tilde{\Psi}$ represents the set $\{\bar{\phi}_I, \bar{\psi}_{O}\}$ and ${\Psi}$ denotes $\{\phi_I,\psi_O\}$.
The non-linear contribution to the action is $\mathcal{S}_{int}[\tilde{\Psi},\Psi]=\int_w \bar{\psi}_O[\frac{c_2}{2}\delta \phi_I(w_1)\delta \phi_I(w-w_1)]+\cdots$.
Physical quantities can be expressed in terms of correlation functions of fields $\Psi$ and $\tilde{\Psi}$, taken with the statistical weight $e^{-\mathcal{S}[\tilde{\Psi},\Psi]}$,
\begin{equation}\label{8}
\langle\Psi \tilde{\Psi}\rangle=\frac{\int \mathcal{D}[i\tilde{\Psi}]\int \mathcal{D}[\Psi] \Psi \tilde{\Psi}e^{-\mathcal{S}[\tilde{\Psi},\Psi]}}{\int \mathcal{D}[i\tilde{\Psi}]\int \mathcal{D}[\Psi] e^{-\mathcal{S}[\tilde{\Psi},\Psi]}}.
\end{equation}
In order to compute the correlation function involving response fields, it is useful to introduce the generating functional,
\begin{equation}\label{9}
\mathcal{Z}[\tilde{J},J]=\langle\exp \int_t \sum_\alpha (\tilde{J}_\alpha(t) \tilde{\Psi}_\alpha(t)+J_\alpha(t) \Psi_\alpha(t))\rangle
\end{equation}
where $\alpha$ represents the set $\{\phi_I,\psi_O\}$,
for which the required correlation functions are obtained via functional derivatives of $\mathcal{Z}$ with respect to the appropriate source fields.

The procedure is readily implemented for the Gaussian theory with statistical weight 
$e^{-\mathcal{S}_0[\tilde{\Psi},\Psi]}$. In Fourier space, we can write the harmonic function as,
\begin{equation}\label{11}
\mathcal{S}_0[\tilde{\Psi},\Psi]= \int_w \sum_\alpha  
\begin{pmatrix}
\tilde{\Psi}_\alpha(-w) & {\Psi}_\alpha (-w)
\end{pmatrix} \mathcal{M} \begin{pmatrix}
\tilde{\Psi}_\alpha(w) \\ {\Psi}_\alpha (w)
\end{pmatrix}
\end{equation}
with the Hermitian coupling, a (4,4) matrix $\mathcal{M}(w)$.
With the aid of Gaussian integrals, we obtain,
\begin{equation}\label{12}
\mathcal{Z}_0[\tilde{J},J]=\exp \left[\frac{1}{2} \int_w \sum_\alpha  
\begin{pmatrix}
\tilde{J}_\alpha(-w) & {J}_\alpha (-w)
\end{pmatrix} \mathcal{M}^{-1} \begin{pmatrix}
\tilde{J}_\alpha(w) \\ {J}_\alpha (w)
\end{pmatrix}\right].
\end{equation}
From Eq.(\ref{12}), we now directly infer the matrix of two point correlation functions in the Gaussian ensemble with the inverse of harmonic coupling matrix $\mathcal{M}$. 

\section{\bf Applications:} 
As a first application we apply the field-theoretic formalism to the
push-pull network, which can be exactly solved for the error
(Eq.(\ref{optimal})). In the process we illustrate the way the
diagrammatic expansion works in the context of signaling networks,
making it possible to apply the theory to more complicated systems.
\paragraph{ \bf A. Push-Pull network:}
The calculation of the error (Eq.(\ref{optimal})) in terms of the
control variable (the average number of phosphatase molecules per cell
($\bar{P}$)) requires the correlation functions $\langle\delta O
\delta I\rangle$, $\langle\delta O^2\rangle$ and $\langle\delta
I^2\rangle$. These can be expressed in terms of the matrix elements of
$\left( \mathcal{M}^{-1} \right)_{mn}$ (Eq.(\ref{11})). Subscripts $m
$ and $n$ represent the $m^{th}$ row and $n^{th}$ column,
respectively. For example, $\left( \mathcal{M}^{-1} \right)_{33}$ is
the correlation function
$\langle\delta\phi_I(-w)\delta\phi_I(w)\rangle$. Similarly we can
obtain other correlation functions.  Now we can compute, power spectra
for the input and output molecules by evaluating the correlation
functions of kinase and substrate populations by using
Eq.~(\ref{12}). We use perturbation theory for the action
corresponding to the push-pull network to compute the non-linear
contribution to the correlation function.

We obtain the following expressions for the power spectra,

\begin{eqnarray}\label{correlations}
\langle\delta\phi_I(-w) \delta \phi_I(w)\rangle_0&=& \frac{2 \gamma_I \langle\phi_I\rangle}{(-iw+\gamma_I)(iw+\gamma_I)}\nonumber \\
\langle\delta\psi_O (-w) \delta \psi_O (w)\rangle_0&=&\frac{c_1^2 2 \gamma_I \langle\phi_I\rangle}{(-iw+\gamma_O)(-iw+\gamma_I)(iw+\gamma_O)(iw+\gamma_I)} \nonumber \\
&&+
\frac{2 \gamma_O \langle\phi_O\rangle}{(-iw+\gamma_O)(iw+\gamma_O)}\nonumber \\
\langle\delta\phi_I(-w) \delta \psi_O(w)\rangle_0&=& \frac{2 c_1 \gamma_I \langle\phi_I\rangle}{(-iw+\gamma_O)(-iw+\gamma_I)(iw+\gamma_I)}
\end{eqnarray}
The $\langle\cdots\rangle_0$ is taken with respect to the non-interacting theory ($S_{int}[\tilde{\Psi},\Psi]=0$ in Eq.(\ref{action})).
Using these functions, the error ($E$) and gain ($G$) are given by,

\begin{equation}\label{error}
E=\frac{\langle(\delta \psi_O -G \delta\phi_I)^2\rangle}{G^2\langle\delta \phi_I^2\rangle},  \  \  \  \ \  \ G=\frac{\langle{\delta \psi_O}^2\rangle}{\langle\delta\phi_I\delta \psi_O \rangle} .
\end{equation}
By  inserting the expressions for the correlation functions in Eq.(\ref{correlations}) into Eq.(\ref{error}), and integrating over $w$, we obtain the minimum relative error for the linear push-pull network,

\begin{equation}\label{linear}
E=1-\frac{\bar{I} \gamma_O^2 \sigma_1^2}{(\gamma_I+\gamma_O)^2}\left[ \gamma_O \sigma_0 + \sigma_1^2 \frac{\gamma_O \bar{I}}{\gamma_O+\gamma_I}\right]^{-1}.
\end{equation}
Higher order corrections to the power spectra $\langle\delta\psi_O
(-w) \delta \psi_O (w)\rangle$ are calculated using perturbation
theory by evaluating the Feynman diagrams (Fig.(\ref{fig:Feynman})),

\begin{equation}
\langle\delta \psi_O \delta \psi_O\rangle=\frac{\langle\delta \psi_O
  \delta \psi_O \sum_l^\infty
  (-\mathcal{S}_{int}[\tilde{\Psi},\Psi])^l /l!\rangle_0}{\langle
  \sum_l^\infty (-\mathcal{S}_{int}[\tilde{\Psi},\Psi])^l
  /l!\rangle_0}.
\end{equation}
For example, the second order contribution to the $\langle\delta\psi_O (-w) \delta \psi_O (w)\rangle$ arising from the loop in Fig.(\ref{fig:Feynman}) is $\Omega_2^2 \frac{2!  \bar{I}^2}{\gamma_O(\gamma_O+2\gamma_I)}$~(see Appendix A for details). 
The coefficient $\Omega_2^2$ is given by $\Omega_2^2=\frac{c_2^2}{4}+\frac{c_3^2}{4}+\frac{\bar{I}}{4}c_2 c_4 +\cdots$. {Higher
order terms have a similar structure:}  for example, the third order
contribution to the power spectra is $\Omega_3^2 \frac{3!
  \bar{I}^2}{\gamma_O(\gamma_O+3\gamma_I)}$, with
$\Omega_3^2=\frac{c_3^2}{36}+\frac{c_4^2}{16}+\frac{\bar{I}}{36}c_3
c_5 +\cdots$.  By evaluating all the diagrams in
Fig.(\ref{fig:Feynman}), we obtain the final expression for the
relative error,
\begin{equation}\label{final}
E=1-\frac{\bar{I} \gamma_O^2 \sigma_1^2}{(\gamma_I+\gamma_O)^2}\left[ \gamma_O \sigma_0 +\sum_{n=1}^{\infty} \Omega_n^2 \frac{n!\gamma_O \bar{I}^n}{\gamma_O+n\gamma_I}\right]^{-1}, ~ ~ \text{with} ~ \Omega_1=\sigma_1.
\end{equation}

The form of the result in Eq.(\ref{final}) coincides with the exact
expression (Eq.(\ref{exact})) for the relative error previously
obtained~\cite{MH14PRX} by using an entirely different approach based
on umbral calculus.~{However, the coefficients $\Omega_n$ are
  expressed in terms of the coefficients $c_n$ used in the series for
  $R(I)$ (Eq.(\ref{rphi})) rather than $\sigma_n$.  The two kinds of
  coefficients are non-trivially related through,
\begin{equation}
  \sigma_n= \sum\limits_{m=0}^{\infty} \sum\limits_{p=0}^{m} {\sum\limits_{q=0}^{p}} \frac{1}{m!} \left(\begin{matrix} m\\p  \end{matrix} \right)\left(\begin{matrix} q\\n  \end{matrix} \right) S_{pq} (-\bar{I})^{m-p}\bar{I}^{q-n}c_m,
\end{equation}
where $S_{pq}$ are Stirling's numbers of second kind.  
For all $n$, the leading order term $\frac{c_n^2}{n!^2}$ of $\Omega_n$
is the same as the leading order term of $\sigma_n$.}
%making Eq.(\ref{final}) an excellent approximation to the exact result Eq.(\ref{exact})

The sum within the bracket in Eq.(\ref{final}) is composed of
non-negative terms. The minimal sum $E$ is obtained by setting
$\Omega_n =0$ for all $n\ge 2$. Thus, $E$ is bounded from below by
$E\ge 1-\frac{\bar{I} \gamma_O^2
  \sigma_1^2}{(\gamma_I+\gamma_O)^2}\left[ \gamma_O \sigma_0 +
  \sigma_1^2 \frac{\gamma_O
    \bar{I}}{\gamma_O+\gamma_I}\right]^{-1}$. The term on the right
hand side is minimized with respect to $\gamma_o$ when $\gamma_o=
\gamma_I \sqrt{1+\tilde{\Lambda}}$, with
$\tilde{\Lambda}=\bar{I}\sigma_1^2/\sigma_0 \gamma_I$. At the optimal
$\gamma_O$, the equality becomes
$E=2/(1+\sqrt{1+\tilde{\Lambda}})\equiv E_{opt}$.  As $\sigma_1 $
increases, $\tilde{\Lambda}$ becomes large which is desirable for high
fidelity signal transduction. As long as $R(I)$ is approximately
linear in the vicinity of $\bar{I}$, the corrections $\sigma_n$ (or
$\Omega_n$) for $n >2$ are negligible, and $E$ is close to
$E_{opt}$. The coefficients $\sigma_n$ for $n > 2$ must be
non-negligible when $\sigma_1$ is sufficiently large. Such a highly
sigmoidal input-output response, known as
ultra-sensitivity~\cite{Goldbeter81PNAS}, is biologically realizable
in certain regimes of signaling cascades. {In the limit of a nearly
  step-like response,} non-linearity in $R(I)$ becomes appreciable
around $\bar{I}$, distorting the output signal and leading to $E$ that
is larger than $E_{opt}$. Because $E$ increases with
$\tilde{\Lambda}$ in this {limit}, the benefits of ultra-sensitivity vanish.

\paragraph{ \bf B. Signaling Cascades:}
A natural extension is to consider a cascade created by an array of
connected push-pull networks.  Indeed, in some biological signaling
pathways external perturbation is transmitted through a cascade of
reactions involving successive activation by kinases and deactivation
by phosphatases. An example is the stimulation of a {receptor}
tyrosin kinase by epidermal growth factor, {which results in downstream
responses of the MAPK network~\cite{Heinrich02MC,Schoeberl02Nature}}.

Because sections $B$, $C$ and $D$ are related, we explain briefly the results
in order to ensure that the relationship between these sections are clear. In this section we  describe
 the two cascade network using the field theory framework, and the coarse-graining 
procedure needed for obtaining an analytic expression for optimal error. 
In section $C$, we show that the two cascade network behaves as noise filter with a time delay, $\alpha^{-1}$. 
By mapping the cascade to a push-pull network with an intermediate, we show in section $D$ that
$\alpha$ can be exactly calculated. Thus, the results in the three sections provide an analytic 
theory for optimal signaling in the two cascade network.

Consider a two step series enzymatic cascade (Fig.(\ref{fig:cas}))
modeled as a sequence of two enzymatic push-pull loops stimulated by
an upstream enzyme. In the first loop, an upstream enzyme, $K$
phosphorylates the substrate, $S$, to produce $S^*$, converting it
from an inactive to active state. Phosphatase ($P$) dephosphorylates
$S^*$ to an inactive state $S$. In the second loop, $S^*$ acts as the
enzyme for the phosphorylation of $T$ and $P$, the corresponding
phosphatases.  The series of chemical reactions involved in this
cascade are,
\begin{eqnarray}\label{cascade}
&&\phi \xrightleftharpoons[\gamma_k]{F} K \nonumber \\
&&K+S\xrightleftharpoons[k_{1u}]{k_{1b}}S_K \xrightarrow[]{k_{1r}} K+S^* ; ~
S^{*}+P\xrightleftharpoons[\rho_{1u}]{\rho_{1b}}S^{*}_{P} \xrightarrow[]{\rho_{2r}} S+P \nonumber \\
&&S^*+T\xrightleftharpoons[k_{2u}]{k_{2b}}S^*_T \xrightarrow[]{k_{2r}} S^*+T^* ;~T^{*}+P\xrightleftharpoons[\rho_{2u}]{\rho_{2b}}T^{*}_{P} \xrightarrow[]{\rho_{2r}} T+P 
\end{eqnarray}
where $S_K$, $S^{*}_{P}$, $S^*_T$ and $T^{*}_{P}$ are the reaction
intermediates, and $ k_{ib}$, $k_iu$, $k_{ir}$, $ \rho_{ib}$,
$\rho_iu$ and $\rho_{ir}$ , $i=1,2$, are the rate constants of the
stochastic biochemical reactions in the cascade. The input signal
$K+S_K$ is transduced into the active substrate output $T^*+T_P^*$.
In an insightful article~\cite{Heinrich02MC}, a deterministic approach
was used to analyze the system of chemical reactions in
Eq.(\ref{cascade}). Here we assume that the reactions are
stochastic. In order to develop analytical results we only consider
fluctuations of all species that deviate linearly from their mean
values. The validity of the asumption is established by comparing the
results with kinetic Monte Carlo (KMC) simulations.

For the network in Fig.(\ref{fig:cas}), the procedure outlined earlier leads to a Schr\"{o}dinger-like equation with the following Hamiltonian,
\begin{eqnarray}
\mathcal{H}=&&-F(\bar{K}-1)-\gamma_k (1-\bar{K})K-k_{1b}(\bar{S}_K-\bar{K}\bar{S})KS -k_{1u} (\bar{K} \bar{S}-\bar{S}_K) S_K\nonumber \\&&-k_{1r}(\bar{S}^*\bar{K}-\bar{S}_K)S_K-\rho_{1u}(\bar{S}^*\bar{P}-\bar{S}_P^*)S_P^*
-\rho_{1b}(\bar{S}_P^*-\bar{S}^*\bar{P})S^* P\nonumber \\&&-\rho_{1r}(\bar{P}\bar{S}-\bar{S}_P^*)S_P^*  
-k_{2b}(\bar{S}_T^*-\bar{S^*}\bar{T})T S^*-k_{2u} (\bar{S^*} \bar{T}-\bar{S}_T^*) S_T^*\nonumber \\&&-k_{2r}(\bar{S}^*\bar{T}^*-\bar{S}_T^*)S_T^*-\rho_{2u}(\bar{T}^*\bar{P}-\bar{T}_P^*)T_P^*-\rho_{2b}(\bar{T}_P^*-\bar{T}^*\bar{P})T^*_P-\rho_{2r}(\bar{P}\bar{T}-\bar{T}_P^*)T_P^*. \nonumber \\
\end{eqnarray} 

{We can approximately map the two-step cascade into a two-species
  coarse-grained network, which acts like a noise filter, as described
  in detail in Ref.~\cite{MH14PRX}. Consider a signaling pathway
  (Fig.(\ref{fig:push-pull})) with time varying input $I(t)$ and time
  varying output $O(t)$. These are the total populations (free and
  bound) of the input and output active kinases, with $I = K + S_K$
  and $O = T^* + T_P^*$. The} upstream pathway provides an effective
production rate $F$ of input $I$, while the output $O$ results from
the reaction $I \xrightarrow[]{R(I)} I+O$. As before, $\gamma_I$ and
$\gamma_O$ are the degradation rates for the input and output
respectively, mimicking the role of phosphatase.  The input and output
correlation functions, evaluated using the field theory formalism,
have the approximate structure,
\begin{eqnarray}\label{6}
P_{ I}(w)&=& \frac{2 F \gamma_{I}^{-2}}{1+(w/\gamma_I)^2} \nonumber \\
P_{ O}(w)&=& \frac{(R_1/\gamma_o G)^2}{1+(w/\gamma_0)^2}\left[G^2 P_{ I}(w)+\frac{2F(G/\gamma_I)^2}{\Lambda} \right] 
\end{eqnarray}
where we have used a linear approximation for $R(I)\approx R_0
\bar{I}+R_1 (I-\bar{I)}$ with $R_0, \ R_1\rangle0$. Optimality is
achieved when $\gamma_O=\gamma_I \sqrt{1+\Lambda}$ with gain
$G=R_1/(\gamma_I(\sqrt{1+\Lambda}-1))$. Relative error with the
minimum $E_{WK}=2/(1+\sqrt{1+\Lambda})$. As before, the fidelity
between output and input is controlled by single dimensionless control
parameter $\Lambda=(R_1/\gamma_I)(R_1/R_0)^2$.  This mapping allows us
to use the general WK result for gain ($G$) and the minimum relative
error ($E_{WK}$) to predict the optimality condition, allowing us to
calculate the minimum possible value of $E$. The results for the error
in terms of the mean number of phosphatase are given by the red lines
in Fig.(\ref{fig:rg1}).

In order to test the accuracy of our theory we simulated the dynamics
of the enzymatic cascade using the KMC method. The relative error $E$
shown in Fig.(\ref{fig:rg1}) is in excellent agreement with the
theoretical predictions. Interestingly, $E$ achieves a minimum at
$\bar{P}=10^5$ molecules/cell, which is ten times larger than the
phosphatase concentration in the one step enzymatic push-pull loop
using similar parameters. Fig.(\ref{fig:rg1}) shows that there is a
well defined narrow range of phosphatase population in which the error
is minimum. The range decreases as $\Lambda$ decreases
(Fig.(\ref{fig:rg1})). The minimum value for the relative error does
not reach the value predicted by the WK limit
(Eq.(\ref{optimal})). {As we show below, the additional error
  arises from an effective time delay as the signal passes from one
  cascade to another.  We also demonstrate that the time delay can
  alternatively be mimicked by reducing the two cascade system to a
  coarse-grained pathway with an intermediate (Fig.(\ref{fig:cas}b))}.

\paragraph{\bf C. {Noise filtering with time delay:}}
In order to prove that the two-cascade loop {effectively acts like
  a noise filter with time delay}, we derive the condition for minimum
error for the latter following the Bode-Shannon formulation of the WK
theory~\cite{Bode50PIRE}.  In this scenario, the transmitted signal can
only be recovered after a constant delay, $\alpha$.  The output
$O(t)$ is produced by convolving the corrupted signal (input
$G I(t)$+ noise $n(t)$) with a causal filter $H(t)$. In Fourier space,
we obtain,
\begin{equation}\label{convolution}
O(w)=H(w) c(w) = H(w) (G I(w)+n(w))
\end{equation}
where $x(w)=\int_{-\infty}^{\infty} dw~x(t) e^{-iwt}$ for the time
series $x(t)$.  The relative error is given by~\cite{Bode50PIRE},
\begin{equation}\label{errorF}
E=\frac{\int_{-\infty}^{\infty} \frac{dw}{2\pi} [| H(w) |^2 P_n(w) +| H(w)-1 |^2 P_I(w)]}{\int_{-\infty}^{\infty} \frac{dw}{2\pi} P_I(w)}
\end{equation}
where $P_I(w)$ and $P_n(w)$ are the power spectral densities (PSDs) of
$G I(t)$ and $n(t)$ respectively.  We need to minimize $E$ in
Eq.(\ref{errorF}) over all possible $H(w)$, with the condition that
$H(t)=0$ for $t < \alpha$. {The} optimal causal filter has the following
form~\cite{Bode50PIRE,Becker15PRL,Hathcock16},
\begin{equation}\label{filterH}
H_{WK}(w)=\frac{e^{i w \alpha }}{P_c^y (w)}\left \{  \frac{P_I(w)e^{-i w \alpha }}{P_c^y(w^*)} \right \}_y.
\end{equation}
The $y$ super and subscript refer to two different decompositions in
the frequency domain.  Causality can be enforced by noting the
following conditions: (i) Any physical PSD, in this case $P_c(w)$
corresponding to the corrupted signal $c(t) = G I(t) + n(t)$, can be
written as $P_c(w) = |P_c^y(w)|^2$. The factor $P_c^y(w)$, if treated
as a function in the complex $w$ plane, does not have zeros and poles
in the upper half-plane ($\text{Im}~w\rangle 0$). (ii) We also define
an additive decomposition denoted by $\{ F(w)\}_y$ for any function
$F(w)$, which consists of all terms in the partial fraction expansion
of $F(w)$ with no poles in the upper half-plane. By using the PSDs,
$P_I(w)=\frac{2 G^2 \gamma_I \bar{I}}{w^2+\gamma_I^2}$ and
$P_c(w)=\frac{2 G^2 \gamma_I \bar{I}}{w^2+\gamma_I^2}+\frac{2 G^2
}{\gamma_I \Lambda}$, we obtain the following optimal filter
$H_{WK}(w)$,
\begin{equation}
H_{WK}(w)=\frac{e^{\alpha(iw-\gamma_I)}\gamma_I(\sqrt{1+\Lambda}-1)}{\gamma_I(\sqrt{1+\Lambda}-iw)}.
\end{equation}
In the limit $\alpha\ll \gamma_I^{-1}$, the optimal error $E_{WK}$
takes the following form~\cite{Hathcock16},
\begin{equation}\label{delay}
E_{WK}=\frac{2}{1+\sqrt{1+\Lambda}}+\frac{2\Lambda \alpha
  \gamma_I}{(1+\sqrt{1+\Lambda})^2},
\end{equation}
where second term in the above equation is the correction due to the
time delay to the WK minimum value of the relative error for an
instantaneous filter ($\alpha \rightarrow 0$).  The correction is
positive for all values of $\alpha$ and $\Lambda$, which implies that
time delay must increase the error in signal transmission.~If we add
this correction to the WK minimum result for the relative error of
instantaneous filter (Eq.(\ref{optimal})), for specific values of
$\alpha$ calculated explicitly in the following section, we recover the minimum relative error in the signaling
cascade. Thus, the two step enzymatic cascade minimizes the noise but
behaves like a single step network with a time delayed filter.

\paragraph{\bf D. {Deriving the time delay $\alpha$ by mapping onto a three-species pathway with an intermediate:}}
{Alternatively, we can derive an explicit expression for the delay
  parameter $\alpha$ by using a different mapping for the original
  cascade.  Instead of mapping onto a two-species network of $I$ and
  $O$ with a time delay, we map onto a three-species network
  (Fig.(\ref{fig:cas}b)) with $I$, $M$, and $O$.  Here there is no
  explicit time delay, but an additional species $M$ that will play
  the role of a ``pseudo'' intermediate mimicking the effect of the
  time delay.}  This network is governed by the reactions: $\phi
\xrightarrow[]{F} I$, $I \xrightarrow[]{R_a(I)} I+M$, $M
\xrightarrow[]{R_b(M)} M+O$, $I\xrightarrow[]{\gamma_I}\phi$,
$M\xrightarrow[]{\gamma_M}\phi$ and $O\xrightarrow[]{\gamma_O}\phi$.
The production functions have the linear form:
$R_a(I)=\sigma_{a0}+\sigma_{a1}(I-\bar{I})$ and
$R_b(M)=\sigma_{b0}+\sigma_{b1}(M-\bar{M})$.  Earlier analysis of this
network~\cite{Hathcock16} has shown that it behaves like a time
delayed filter, with the minimal error in the same form as
Eq.(\ref{delay}), with $\alpha=\gamma_M^{-1}$ and effective
$\Lambda=\Lambda_b \sqrt{1+\Lambda_a}$, where
$\Lambda_a=\bar{I}\sigma_{a1}^2/\sigma_{ao} \gamma_I$ and
$\Lambda_b=\bar{M}\sigma_{b1}^2/\sigma_{bo} \gamma_M$.

The {original} signaling cascade (Fig.(\ref{fig:cas}a)) can be
mapped onto the three-species pathway (Fig.(\ref{fig:cas}b)).
{This involves identifying the population $S^* + S_P^* = M$ as a
  ``pseudo'' intermediate, with an effective degradation
  $\gamma_M$. The mapping can be carried out by comparing PSDs between
  the two models.  For the three-species network these are given by,}
\begin{eqnarray}\label{psd1}
&&P_{\delta I}(w)=\frac{2 \gamma_I \bar{I}}{w^2+\gamma_I^2} \\ \nonumber
&& P_{\delta O}(w)=\frac{G^{-2}\sigma_{a1}^2\sigma_{b1}^2}{(w^2+\gamma_M^2)(w^2+\gamma_O^2)}\left( G^2  P_{\delta I}(w) +\frac{2G^2 \gamma_M \bar{M} }{\sigma_{a1}^2}+\frac{G^2 (w^2+\gamma_M^2)2\gamma_O \bar{O}}{\sigma_{a1}^2 \sigma_{b1}^2} \right)
\end{eqnarray}
 Now, the PSDs for signaling cascade calculated from Doi-Peliti formalism are given by
 \begin{equation}\label{psd2}
 P_{\delta I}(w)=\frac{\sum_{i=0}^{N-1} n_{\delta I,i} w^{2i}}{1+\sum_{i=1}^{N} d_{\delta I,i} w^{2i}}~ \text{and}~P_{\delta O}(w)=\frac{\sum_{i=0}^{N-1} n_{\delta O,i} w^{2i}}{1+\sum_{i=1}^N d_{\delta O,i} w^{2i}}
 \end{equation}
where the $w$-independent parameters $n_{\delta I,i}$, $n_{\delta O,i}$, $d_{\delta I,i}$  and $d_{\delta O,i}$ are related to the rate coefficients in the cascade reactions (Eq.(\ref{cascade})). Here, $N=7$ corresponds to the number of independent dynamical variables ($K,S_K,S^*,S_P^*,T,T_P^* ~\text{and}~ T^*$). By mapping Eq.(\ref{psd2}) into Eq.({\ref{psd1}}), we can extract the degradation rate of intermediate species ($S^* + S_P^*$), $\gamma_M$ in terms of coefficients in Eq.(\ref{psd2}), 
\begin{equation}\label{gammaM}
\gamma_M^2=\frac{1}{2}[A+\sqrt{(A^2+4B)}],
\end{equation}
with $A=\frac{d_{\delta O, 2}}{d_{\delta O, 3}} -\gamma_I^2$ and $B=A
\gamma_I^2-\frac{d_{\delta O, 1}}{d_{\delta O, 3}}$.  The time delay
parameter $\alpha=\gamma_M^{-1}$ in the signaling cascade. With
{this identification for $\alpha$} we have a complete theory for
$E$, with no adjustable parameter, as a function of the control
parameter, the mean phosphatase levels.  It is tempting to speculate
that a multiple ($>2$) step cascade {might also be} mathematically
equivalent to a network with a single pseudo intermediate.

\paragraph{ \bf E. Enzymatic Push-Pull Loop:}
{In considering the cascade model, we focused on the case where
  fluctuations around mean populations levels were small enough that
  the linear approximation is valid.  To study the effects of
  non-linearity, we will look at a simpler system (one stage of the
  cascade) but without any constraints on the size of the
  fluctuations.}  A microscopic model for the enzymatic push-pull
network is shown in Fig.(\ref{fig:enzymatic}).  The upstream enzyme,
$K$ phosphorylates a substrate $S$ to $S^*$, thereby converting it
from an inactive to an active state. The effective production rate in
the upstream pathway for enzyme $K$ is $F$. The degradation rate for
$K$ is $\gamma_K$. The enzyme is either free ($K$) or bound to
substrate ($S_K$). The input $I$ is the total enzyme population
$I=K+S_K$. Phosphatase, $P$, on the other hand dephosphorylates the
active substrate $S^*$ to an inactive state $S$.  The output of the
two phosphorylation cycle is $O=S^{*}+S_P^{*}$.

The biochemical reactions for the enzymatic network with the corresponding rate constants are,
%\section{Reactions:}
\begin{eqnarray}\label{1}
&&\phi \xrightleftharpoons[\gamma_k]{F} K \nonumber \\
&&K+S\xrightleftharpoons[k_{u}]{k_{b}}S_K \xrightarrow[]{k_{r}} K+S^* \nonumber \\
&&S^{*}+P\xrightleftharpoons[\rho_{u}]{\rho_{b}}S^{*}_{P} \xrightarrow[]{\rho_{r}} S+P .
\end{eqnarray}
In the stochastic chemical reactions that govern the
phosphorylation/dephosphorylation steps, the input signal $I= K + S_K$
is transduced into the active substrate output $ S^{*}+ S_P^{*}$.  To
derive the conditions for optimality, we follow the procedure outlined
in the previous section. Starting from the master equation, we can
derive a Schr$\ddot{o}$dinger-like equation with the following
Hamiltonian,
\begin{eqnarray}
\mathcal{H}&=&-F(\bar{K}-1)-\gamma_k (1-\bar{K})K-k_{b}(\bar{S}_K-\bar{K}\bar{S})KS \nonumber \\
&&-k_{u} (\bar{K} \bar{S}-\bar{S}_K) S_K-k_{r}(\bar{S}^*\bar{K}-\bar{S}_K)S_K-\rho_{u}(\bar{S}^*\bar{P}-\bar{S}_P^*)S_P^*\nonumber \\
&&-\rho_{b}(\bar{S}_P^*-\bar{S}^*\bar{P})S^*_P-\rho_{r}(\bar{P}\bar{S}-\bar{S}_P^*)S_P^* 
\end{eqnarray} 
The field variables $\bar{\phi}$ are associated with creation
operators of corresponding population. Similarly $\phi $ correspond to
annihilation operators.  After using coherent-state path integral
formalism, we arrive at the expression for the action corresponding to
the enzymatic push-pull loop
from which we calculate the power spectra for the input and output.   

As in the signaling cascade network described in the previous section,
we approximately map the complete enzymatic network into a noise
filter~\cite{MH14PRX}. The input and output correlation functions,
evaluated using field theory formalism, have the approximate structure
given in Eq.(\ref{6}).  Starting from the full dynamical equations
(Eq.(\ref{1})), we compute correlation functions using field theory by
solving the Wiener-Hopf relation in Eq.(\ref{WK}),
%\begin{equation}\label{6a}
%C_{cs}(t)=\int_{-\infty}^t dt' H_{WK}(t-t') C_{cc}(t'), \ \ t\rangle0
%\end{equation}
%Where $C_{xy}(t)\equiv \langlex(t')y(t'+t)\rangle$ is the two point correlation in time series $x$ and $y$ with time difference $t$. Subscript $c$ corresponds to the corrupted signal $c(t)=s(t)+\eta(t)$, with input signal $s(t)=G  I(t)$.
for the optimal function $H_{WK}(t)$.

Correlation functions of input and output calculated for enzymatic
push-pull loop have the approximate form of Eq.(\ref{6}), with
effective values of parameters $\gamma_I$, $\gamma_O$, $R_1$ and
$\Lambda$ which have been expressed in terms of loop reaction rate
parameters. This mapping allows us to use WK result for the gain ($G$)
and minimum relative error ($E_{WK}$) to predict the optimality and
minimum possible value of $E$. The results for the error in terms of
the mean number of phosphatase are given by the solid lines in
Fig.(\ref{fig:error}).

%\paragraph{Results:}

In order to illustrate the accuracy of the theory we performed KMC
simulations by choosing the forward and backward reaction rates in
Eq.(\ref{1}) describing the enzymatic push-pull loop network (all
units are in $s^{-1}$) : $k_{b}=\rho_{b}=10^{-5}$, $k_{u}=0.02$,
$\rho_{u}=0.5$, $k_{r}=3$, $\rho_{r}=0.3$, $F=1$. The deactivation
rate $\gamma_k=0.01 s^{-1}$ of enzyme $K$ which controls the
characteristic time scale over which the input signal varies,
mimicking the role of phosphatase. Mean free substrate and phosphatase
populations are in the ranges $\bar{S}=\bar{P} \sim 10^3-10^5$
molecules/cell.  Fig (\ref{fig:error}) shows that $E$ is a minimum at
a particular value of phosphatase concentration $\bar{P}$, where
optimality condition is satisfied i.e. $\gamma_O=\gamma_I
\sqrt{1+\Lambda}$. For a particular value of $\Lambda=100$, we see
minimum error $E=0.18$ for the enzymatic push-pull loop. The result of
the KMC simulations (purple circles) are in excellent agreement with
the analytical calculation (blue line) for all $\Lambda$ values.

 In the parameter space used in the results in Fig.(\ref{fig:error}),
 a linear theory reproduces the simulation results well.  However,
 deviations from the predictions of the linear theory are expected if
 the input parameters are varied. In order to investigate {these
   deviations} we first obtained the error using the parameter values,
 $k_{b}=\rho_{b}=10^{-3}$, $k_{u}=0.02$, $\rho_{u}=0.5$, $k_{r}=3$,
 $\rho_{r}=0.3$ using KMC simulations. The relative error for
 $\Lambda=100$ is shown in purple line in Fig.(\ref{fig:error1}). The
 blue line, calculated from linear theory predictions, deviates
 substantially from simulations (purple line in
 Fig.(\ref{fig:error1})).  To improve the predictions of the theory we
 calculated second order corrections to $E$. The result, displayed as
 green curve in Fig.(\ref{fig:error1})), shows that there is improved
 agreement between theory and simulations. The non-linear corrections,
 which are substantial, brings the theoretical predictions closer to
 the simulation results, especially near the values of $\bar{S}$ for
 which the error is a minimum (Fig.(\ref{fig:error1}))). We suspect
 that higher order perturbative corrections will further improve the
 results based on the following observation. We fit the dependance of
 the error for ($\bar{S} > 1600$) using the function, $E(\bar{S})= a+b
 (\bar{S}-\bar{S}_{\text{min}})^{1.4}+c(\bar{S}-\bar{S}_{\text{min}})^2$,
 where $a,~b$ and $c$ are constants and $\bar{S}$ is the value of
 $\bar{S}$ at which $E(\bar{S}_{\text{min}})=a$ is a minimum. The
 functional form of $E(\bar{S})$ is the same for the exact simulation
 results, and the predictions of the linear and non-linear theory
 except the coefficients $a,~b$ and $c$ are different.  We, therefore,
 surmise that higher order terms merely renormalize the coefficients,
 keeping unaltered the form of relative error. Consequently, we
 conclude that improved estimates of $a,~b$, $c$ from third and higher
 order contributions should produce predictions in better agreement
 with simulation.

\section{\bf Concluding Remarks:}

In order to assess the accuracy of signal transmission, using the
{mean} square of the error between input and output as a fidelity
measure, we have developed a field theoretic formulation that allows
us to predict conditions for optimal information transfer for an
arbitrary stochastic chemical reaction network. The starting point is
the classical master equation for interacting particle systems, which is mapped 
 to a
  non-Hermitian 'quantum' many-body
Hamiltonian dynamics. Finally, the coherent-state path integral
representation is utilized to arrive at a continuum field theory
description that faithfully incorporates the intrinsic reaction noise
and discreteness of the original stochastic processes. The formulation
allows us to use standard field theory methods to compute the relative
error in the information transfer using perturbation theory to all
orders in non-linearity. This approach leads to an analytical
expression for the minimum relative error in signal transduction. The
usefulness of the general field theory formulation is illustrated
{through} signaling networks of increasing complexity.

Detailed study of an enzymatic push pull loop, the basic unit involved
in complex signaling pathways, {show that it behaves like an
  optimal linear WK noise filter}, as previously established using
entirely different methods~\cite{MH14PRX}. In this particular case,
the joint probability $P(\delta I, \delta O)$ is {approximately
  bivariate Gaussian}, which means the error {$E$ is also directly
  related to the mutual information $M$ in bits between $\delta I$ and
  $\delta O$ as $E = 2^{-2M}$~\cite{Hathcock16}.}

The {two-stage} enzymatic cascade behaves as an optimal filter
without achieving the minimum predicted by the WK theory. We attribute
the deviation to the time delayed response of the cascade. By mapping
the cascade signaling network to a three-species push-pull like model with a
pseudo intermediate state we derived an explicit expression for the
time delay. We show that the time delay is associated with the
degradation rate of the pseudo intermediate state in the
coarse-grained representation of the two-step cascade.  We also
  demonstrate that in those cases where the linear approximation
  breaks down, systematic perturbative corrections can be calculated
  using our theory, which minimize the difference between the findings
  in the simulations and theoretical predictions. The success in this
  example illustrates the power of the formalism.  Analyzing
  experimental data using the framework introduced here will help
  decipher the design principles governing signaling networks in
  biology, and allow us to understand the constraints imposed by noise
  in information transfer.
  
  {\bf Acknowledgements:} We are grateful to the National Science Foundation (CHE 16-61946) for supporting our work. Much of this work was carried out while the authors were in the Institute for Physical Sciences and Technology in the University of Maryland, College Park.

\newpage
\appendix
\section{Appendix A: Second order loop correction to the signaling error for the push-pull network:}
 Here, we illustrate the calculation of $E$ arising from perturbation
 expansion of the field theory for the push-pull network with
 non-linearity explained in the text. To second order the diagram
 needed to compute $E$ is,
\begin{figure}[h]
	\includegraphics[width=0.4\textwidth]{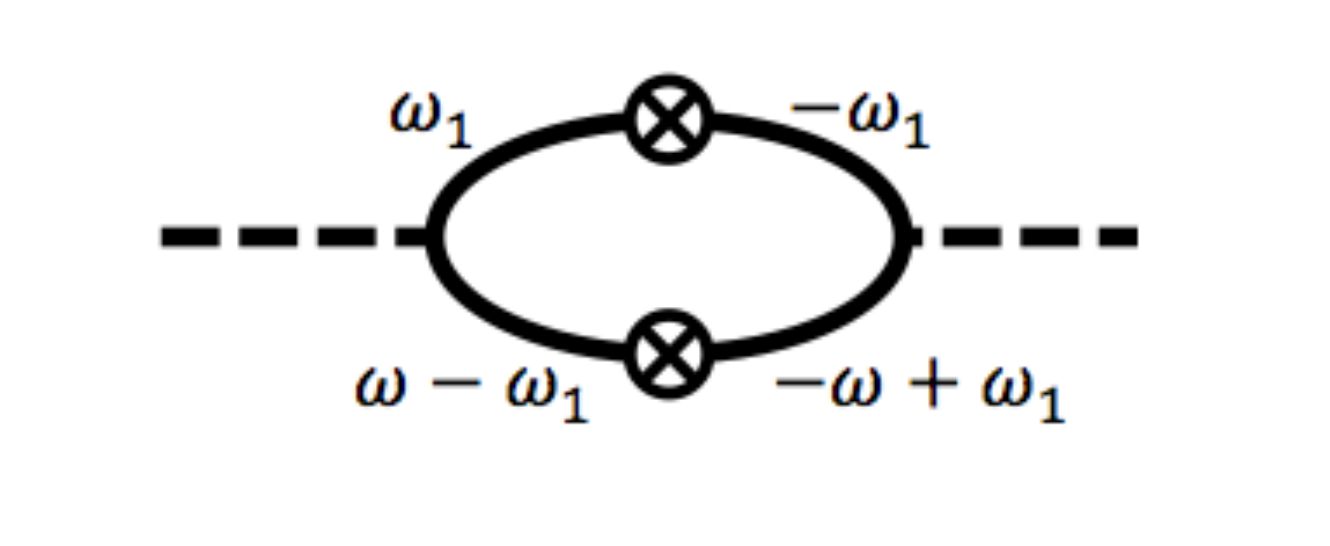}
	%\caption{Feynman Ddiagrams for the correlation function $\langle\delta \psi_O(-w) \delta \psi_O(w)\rangle$. Where, $\sigma_{1}^2=c_{1}(c_{1}+\sum_{m=1}^{n} \frac{(2m+1)!}{2^m}c_{2m+1}\bar{I}^m)$, $\sigma_{2}^2=c_{2}(c_{2}+\sum_{m=2}^{n} \frac{(2m)!}{2^{m-1}}c_{2m}\bar{I}^{m-1})$, $\sigma_{3}^2=c_{3}(c_{3}+\sum_{m=2}^{n} \frac{(2m+1)!}{2^m}c_{2m+1}\bar{I}^m)$ and so on. Evaluation of the diagrams are required in the calculation of the relative errors in the regulatory networks that transmit signals with high fidelity.}.
	\label{fig:Feynman}
\end{figure}
\begin{eqnarray}\label{loop}
E_2&=&2(2 \gamma_I \bar{I})^2 \int\frac{dw}{2\pi} \int\frac{dw_1}{2\pi} \frac{1}{(w_1^2+\gamma_I^2)(w^2+\gamma_O^2) (w_1-w)^2+\gamma_I^2)}\\ \nonumber
&=& 2(2 \gamma_I \bar{I})^2 \int\frac{dw}{2\pi} \frac{1}{(w^2+\gamma_O^2) } \frac{2\pi i}{2\pi}\frac{1}{2i\gamma_I}\left[\frac{1}{w(w-2i\gamma_I)}+\frac{1}{w(w+2i\gamma_I)}\right]\\ \nonumber
&=& 2\frac{(2 \gamma_I \bar{I})^2 }{\gamma_I}\frac{2\pi i}{2\pi}\left[ \frac{1}{2 i \gamma_O} -\frac{1}{4i\gamma_I}\right] \frac{1}{4\gamma_I^2-\gamma_O^2}\\ \nonumber
&=& \frac{2\bar{I}^2}{\gamma_O(\gamma_O+2\gamma_I)}.
\end{eqnarray}
In the first line of the above equation, we perform the complex
integration in the upper half plane by evaluating the residues at
poles $w_1=i\gamma_I$ and $w_1=w+i\gamma_I$, respectively.  Similarly,
in the second line we calculate the residues at poles $w=i\gamma_O$
and $w=2i\gamma_I$.

The coefficient, $\Omega_2^2$ (Eq.(\ref{final})), is diagrammatically
represented as,
\begin{figure}[h]
	\includegraphics[width=0.7\textwidth]{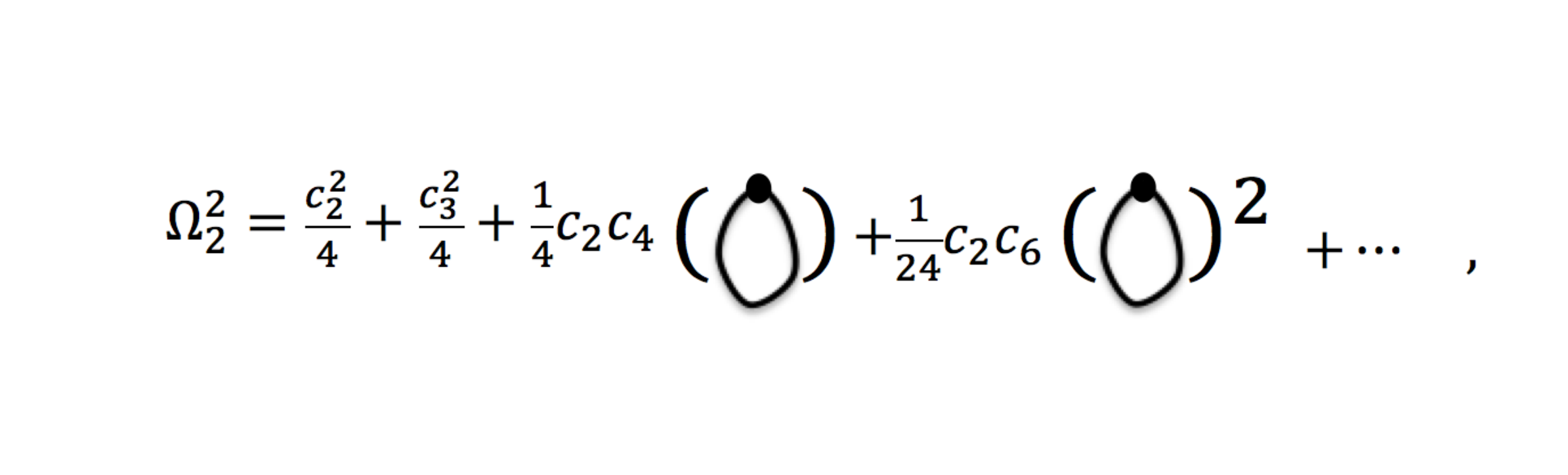}
	%\caption{Feynman Ddiagrams for the correlation function $\langle\delta \psi_O(-w) \delta \psi_O(w)\rangle$. Where, $\sigma_{1}^2=c_{1}(c_{1}+\sum_{m=1}^{n} \frac{(2m+1)!}{2^m}c_{2m+1}\bar{I}^m)$, $\sigma_{2}^2=c_{2}(c_{2}+\sum_{m=2}^{n} \frac{(2m)!}{2^{m-1}}c_{2m}\bar{I}^{m-1})$, $\sigma_{3}^2=c_{3}(c_{3}+\sum_{m=2}^{n} \frac{(2m+1)!}{2^m}c_{2m+1}\bar{I}^m)$ and so on. Evaluation of the diagrams are required in the calculation of the relative errors in the regulatory networks that transmit signals with high fidelity.}
	%\label{fig:Feynman}
\end{figure}\\
where the expression for the loop in the first bracket is $ 2 \gamma_I
\bar{I} \int\frac{dw_1}{2\pi}
\frac{1}{(w_1^2+\gamma_I^2)}=\bar{I}$. The coefficients $\Omega_n$ in
Eq.(\ref{final}) are functions of $c_n$. {In turn, $\Omega_n$s and
$\sigma_n$s are also connected by the relation between $\sigma_n$ and
$c_n$ (see main text). For} all $n$, the leading order
term ($\frac{c_n^2}{n!^2}$) of $\Omega_n$ and $\sigma_n$ {is}
identical.
\newpage

\section {Appendix B: Action for enzymatic push-pull network:}
We give the form of the action here for the enzymatic push-pull
network for which the chemical reaction scheme is given in
Eq.(\ref{1}).~Despite the complexity, the action can be manipulated
using Mathematica in order to obtain general expression for the error.
\begin{eqnarray}\label{actionE}
&&\mathcal{S}[\tilde{\Phi}, \Phi]=\int_w  \bar{\phi}_K[-iw \phi_K-F+\gamma_K \phi_K +k_{b} \phi_K \phi_S-(k_{r}+k_{u})\phi_{S_K}]+ 
\bar{\phi}_{S_K} [-iw S_K-\\ \nonumber&&k_{b} \phi_K \phi_S+(k_{r}+k_{u})\phi_{S_K}]  +
\bar{\phi}_{S^*}[-iw S^*-k_{r} \phi_{S_K}-\rho_{u}\phi_{S_P^*}+\rho_{b}\phi_{S^*} \phi_P]+\bar{\phi}_S [-iw \phi_S+k_{b} \phi_K \phi_S\\ \nonumber &&-\rho_{r} \phi_{S_P^*}-k_{u} \phi_{S_K}]+
\bar{\phi}_{S_P^*}[-iw \phi_{S_P^*}+(\rho_{u}+\rho_{r})\phi_{S_P^*}-
\rho_{b} \phi_{S^*}\phi_P]+\bar{\phi}_P [-iw \phi_P-(\rho_{u}+\rho_{r})\phi_{S_P^*}+\rho_{b} \phi_{S^*}\phi_P]
\\ \nonumber &&+\frac{1}{2} \bar{\phi}_K^2[-F-\gamma_K \phi_K 
 -k_{b} \phi_K \phi_S-(k_{r}+k_{u})\phi_{S_K}]+\frac{1}{2} \bar{\phi}_{S_K}^2[-k_{b} \phi_K \phi_S-(k_{r}+k_{u})\phi_{S_K}]
 +\frac{1}{2} \bar{\phi}_{S^*}^2[-k_{r} \phi_{S_K}\\ \nonumber &&-\rho_{u}\phi_{S_P^*}-\rho_{b}\phi_{S^*} \phi_P]+\frac{1}{2} \bar{\phi}_{S}^2[-k_{b} \phi_K \phi_S-\rho_{r} \phi_{S_P^*}-k_{u} \phi_{S_K}]
 +\frac{1}{2} \bar{\phi}_{S_P^*}^2[-(\rho_{u}+
 \rho_{r})\phi_{S_P^*}-\rho_{b} \phi_{S^*}\phi_P]\\ \nonumber &&+\frac{1}{2} \bar{\phi}_{P}^2[-(\rho_{u}+\rho_{r})\phi_{S_P^*}-\rho_{b} \phi_{S^*}\phi_P-] 
 -k_{b}[\bar{\phi}_{S_K} \bar{\phi}_K-\bar{\phi}_{S_K}\bar{\phi}_S+\bar{\phi}_K \bar{\phi}_S ]\phi_K \phi_S -k_{r}[\bar{\phi}_{S^*}\bar{\phi}_K 
 -\bar{\phi}_{S^*} \bar{\phi}_{S_K}\\ \nonumber &&-\bar{\phi}_K \bar{\phi}_{S_K}] \phi_{S_K} -\rho_{u}[\bar{\phi}_{S^*}\bar{\phi}_P-\bar{\phi}_{S^*}\bar{\phi}_{S_P^*}-\bar{\phi}_P \bar{\phi}_{S_P^*}]\phi_{S_P^*}-\rho_{r}[\bar{\phi}_{S}\bar{\phi}_P-\bar{\phi}_{S^*}\bar{\phi}_{P}-\bar{\phi}_S \bar{\phi}_{S_P^*}]\phi_{S_P^*}-\\ \nonumber &&
 \rho_{b}[-\bar{\phi}_{S^*}\bar{\phi}_{S_P^*}- \bar{\phi}_{P}\bar{\phi}_{S_P^*}+\bar{\phi}_P \bar{\phi}_{S^*}]\phi_{S^*}\phi_P
 -k_{u}[\bar{\phi}_{S}\bar{\phi}_{K}-\bar{\phi}_{K}\bar{\phi}_{S_K}-\bar{\phi}_S \bar{\phi}_{S_K}]\phi_{S_K}.\\ \nonumber &&
\end{eqnarray}

\newpage
%\bibliographystyle{unsrt}
%\bibliography{signalingbib}

\newpage

\begin{figure}
	\includegraphics[width=01.0\textwidth]{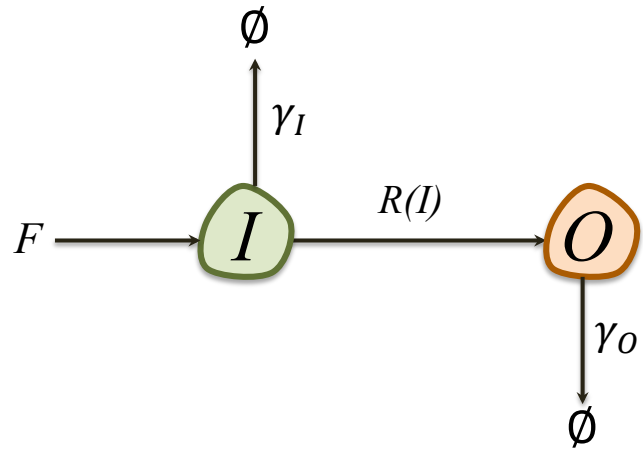}
	\caption{Schematic of a push-pull network, involving an input species $I$ and output species $O$. The production of $O$ from $I$ is controlled by the rate function $R(I)$. The degradation rates for $I$ and $O$ are $\gamma_I$ and $\gamma_O$, respectively.}
	\label{fig:push-pull}
\end{figure}
\begin{figure}
	\includegraphics[width=0.9\textwidth]{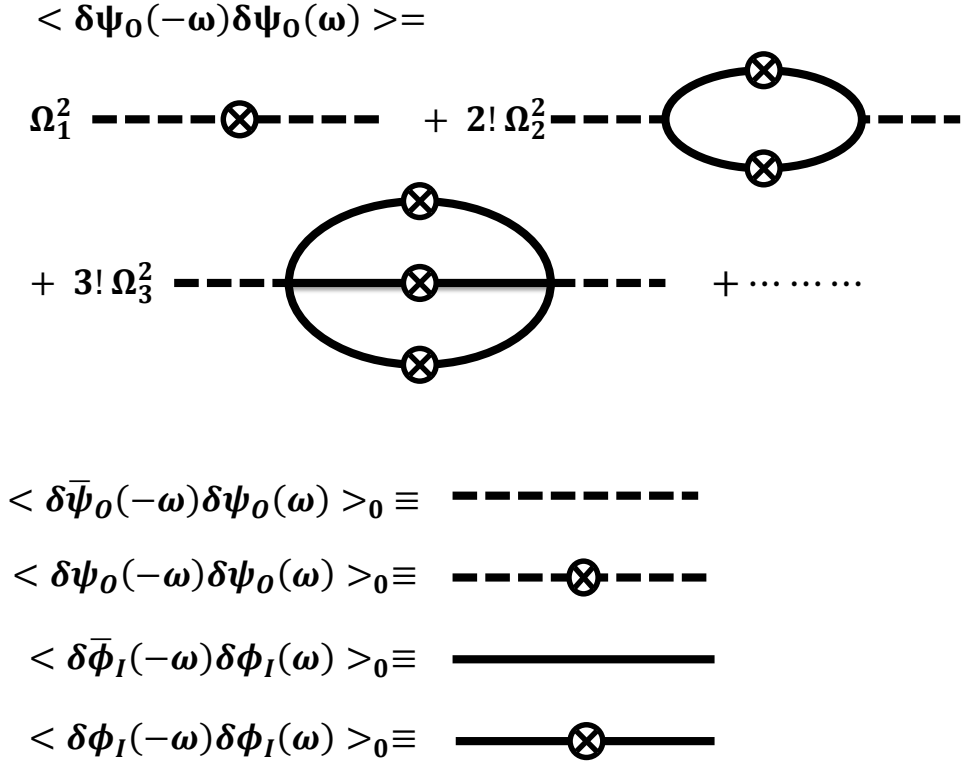}
	\caption{Examples of diagrams for the correlation function $\langle\delta \psi_O(-w) \delta \psi_O(w)\rangle$. The $\Omega_n$s are coefficients with, $\Omega_{1}^2=c_{1}^2$%+\sum_{m=1}^{n} \frac{(2m+1)!}{2^m}c_{2m+1}\bar{I}^m)$
	, $\Omega_2^2=\frac{c_2^2}{4}+\frac{c_3^2}{4}+\frac{\bar{I}}{4}c_2 c_4 +\cdots$, $\Omega_3^2=\frac{c_3^2}{36}+\frac{c_4^2}{16}+\frac{\bar{I}}{36}c_3 c_5 +\cdots$ and so on. }
	\label{fig:Feynman}
\end{figure}

\begin{figure}
	\includegraphics[width=0.6\textwidth]{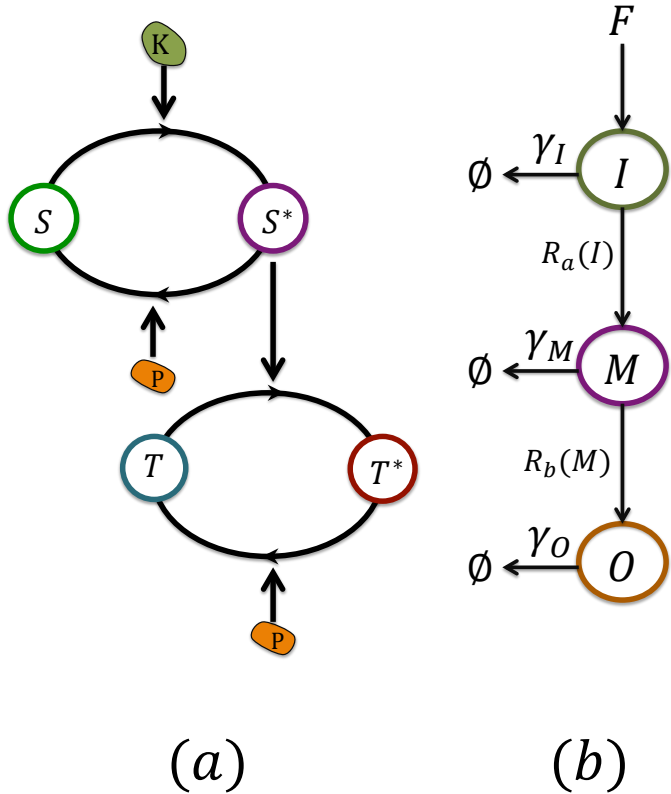}
	\caption{(a) Enzymatic cascade that arises naturally in mitogen activated protein kinase (MAPK) networks. In a caricature of such a network, kinase ($K$) phosphorylates the substrate ($S$), leading to the formation of $S^*$. Deactivation is triggered by reactions with the  phosphatase ($P$). $S^*$ phosphorylates the substrate $T$, producing $T^*$ and $P$ reverts it to the original form through dephosphorylation. The rate parameters in the chemical reactions (Eq.~(\ref{cascade})) used to produce numerical results (in units of $s^{-1}$) are : $k_{1b}=k_{2b}=\rho_{1b}=\rho_{2b}=10^{-5}$, $k_{1u}=0.02$, $k_{2u}=0.3$, $\rho_{1u}=0.5$, $\rho_{2u}=1.0$, $k_{1r}=3$, $k_{2r}=5.0$, $\rho_{1r}=0.3$, $\rho_{2r}=0.1$, 
$F=1$. The deactivation rate $\gamma_k=0.01 s^{-1}$ controlling the characteristic time scale over which the input signal varies. Mean free substrate and phosphatase populations are in the ranges $\bar{S}=\bar{P} \sim 10^4-10^6$ molecules/cell. (b) Three species {coarse-grained} signaling network with the indicated rates is intended to capture the physics of the cascade in (a). The mathematical equivalence between the networks in (a) and (b) is established in the text.}
	\label{fig:cas}
\end{figure}

\begin{figure}
	\includegraphics[width=0.9\textwidth]{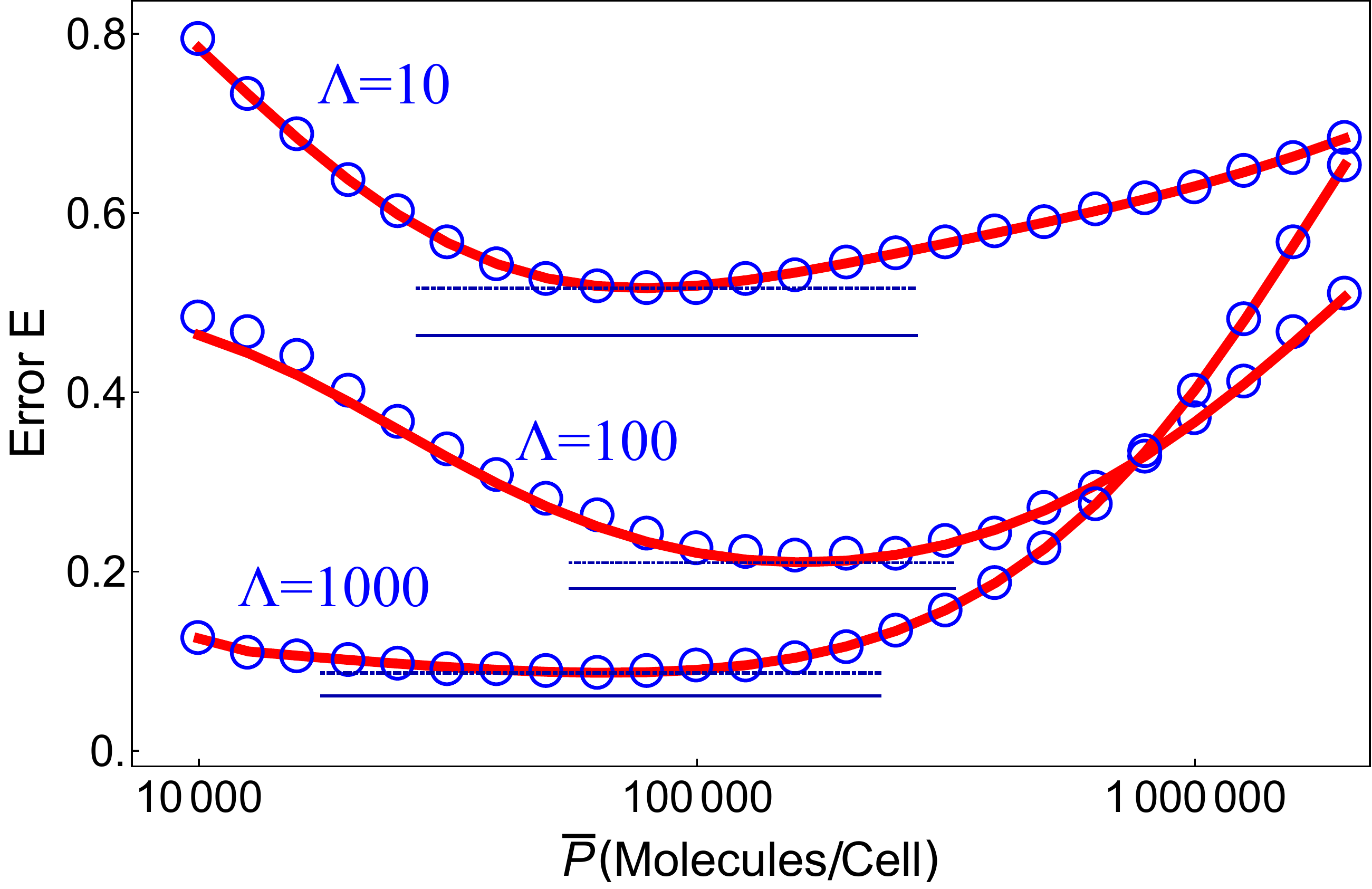}
	\caption{Relative error $E$ for the signaling cascade (red
          lines show the theoretical predictons; blue circles are
          obtained using the kinetic Monte Carlo simulations) for
          three $\Lambda$ values. The blue dashed line gives the
          predictions (Eq.(\ref{delay})) using the WK formalism for
          $E_{WK}$ with time delay. The solid blue line is the minimal
          error corresponding to the {theory without time delay} in
          Eq.(\ref{optimal}). The comparison shows that the two-loop
          cascade behaves as a push-pull network with a time delay. The time delay
          parameter, $\alpha$, is explicitly given in Eq.(\ref{gammaM}). Thus, the theory has no adjustable parameter. }
	\label{fig:rg1}
\end{figure}
\begin{figure}
	\includegraphics[width=0.90\textwidth]{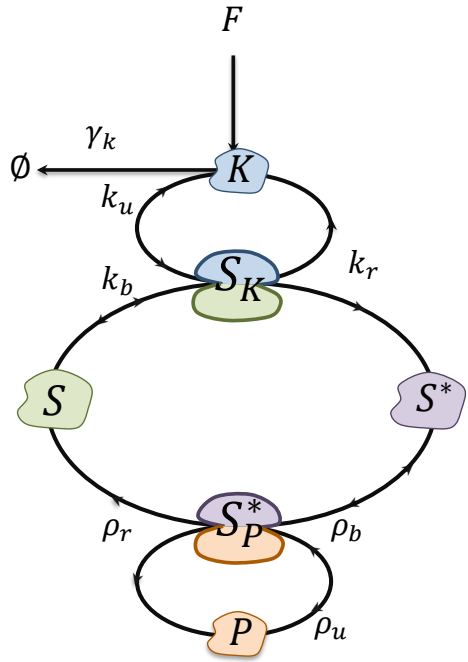}
	\caption{Enzymatic push-pull loop showing phosphorylation of
          the substrate ($S$) by kinase ($K$) to produce the active
          form $S^*$. Phosphatase ($P$) reverts it to the original
          form through dephosphorylation. $S_K$ and $S_P^*$ represent
          the substrate in the complex with kinase and phosphatase,
          respectively. Binding, unbinding and the reaction rate
          constants are shown with arrows.}
	\label{fig:enzymatic}
\end{figure}

\begin{figure}
	\includegraphics[width=0.9\textwidth]{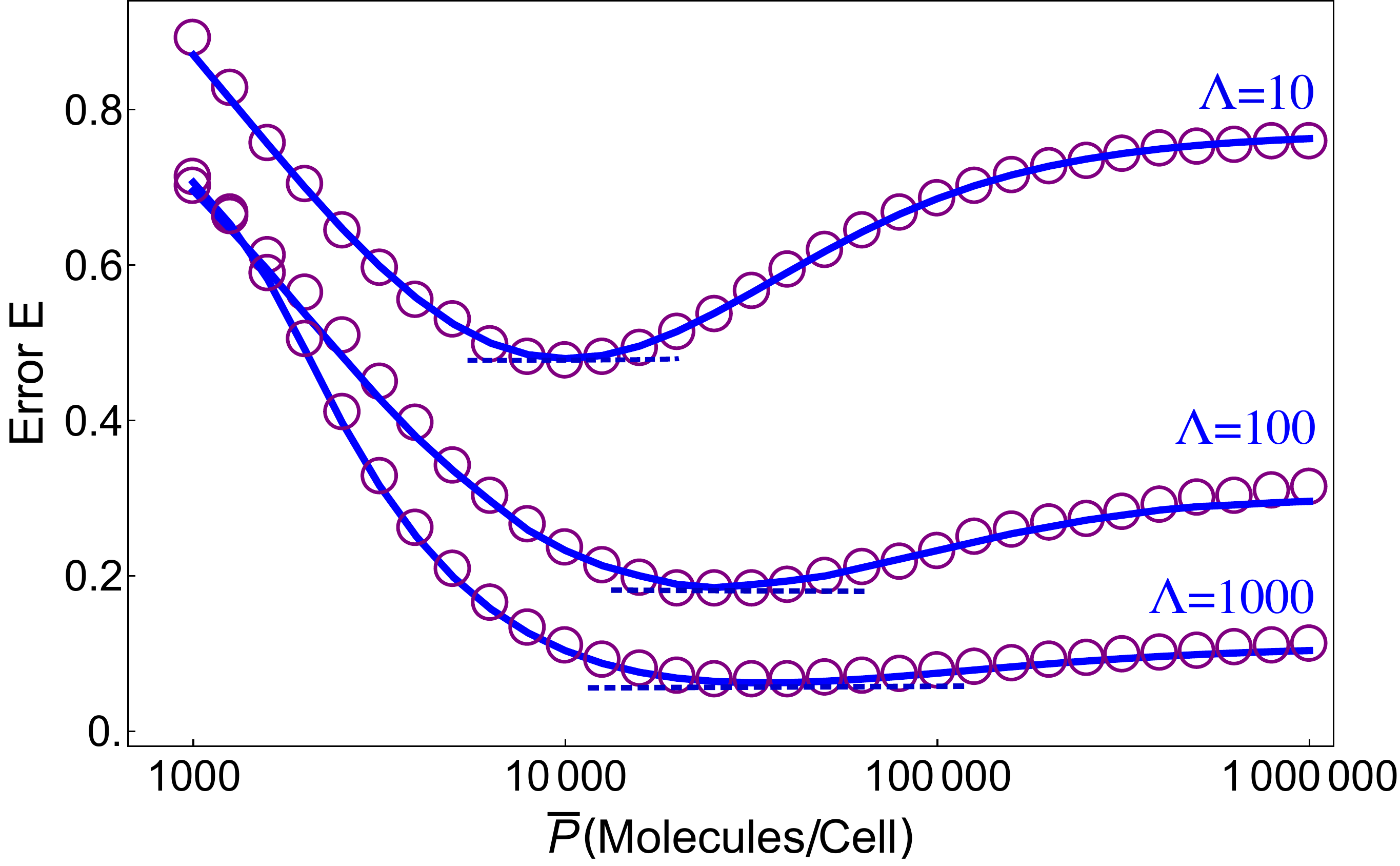}
	\caption{Relative error $E$ for the enzymatic push-pull loop
          {in Fig.(\ref{fig:enzymatic}).} The blue lines correspond
          to theoretical predictions. The KMC simulation results are
          given in purple circles. The dashed line is the minimal
          error corresponding to the WK theory
          (Eq.(\ref{optimal})). The values of the rates corresponding
          to the chemical reactions in the enzymatic push-pull network
          (Eq.(\ref{1})) is given in the main text. For the parameter
          values the predictions of the linear theory are very
          accurate.}
	\label{fig:error}
\end{figure}

\begin{figure}
	\includegraphics[width=0.9\textwidth]{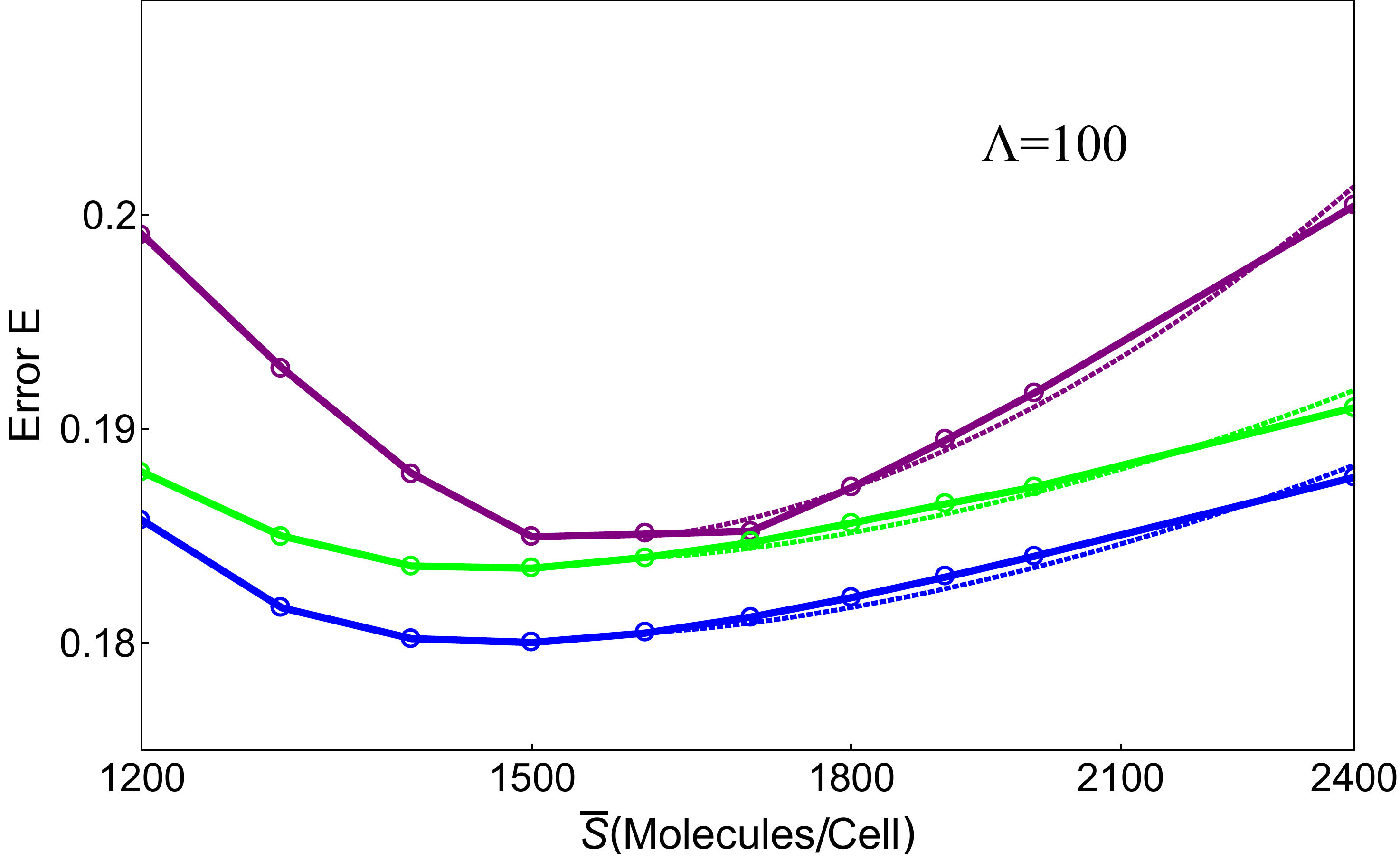}
	\caption{Error $E$ for the enzymatic push-pull loop for
          different values of $\bar{S}$ with $\Lambda=100$. The rate
          parameters used in Eq.(\ref{1}) are
          $k_{b}=\rho_{b}=10^{-3}$, $k_{u}=0.02$, $\rho_{u}=0.5$,
          $k_{r}=3$ and $\rho_{r}=0.3$. The blue line is the result
          calculated using linear theory. The green line results from
          second order corrections to the error $E$. The KMC
          simulation results are given in purple line. Clearly
          inclusion of non-linear corrections improves the predictions
          of the theory in the range of $\bar{S}$ values for which $E$
          is small. Dotted lines are fit with the function
          $a+b(\bar{S}-1600)^{1.4}+c(\bar{S}-1600)^2$ where $a$, $b$,
          $c$ are constants. For all the curves $a$, $b$, $c$ values
          change but the functional form of $E$ as a function of
          $\bar{S}$ is the same. It is likely that if the theory is
          extended beyond second order, there should be further
          improvement by bringing $a$, $b$ and $c$ values closer to
          the simulation results.}
	\label{fig:error1}
\end{figure}

%\begin{figure}
%	\includegraphics[width=0.8\textwidth]{rg5}
%	\caption{Sample trajectories for the scaled input $G\delta I(t)$ (red) and the output $\delta O(t)$ (blue) from KMC simulations of the enzymatic push-pull loop, for $\lambda=100 (S=2.8\times 10^5 )$ and $\bar{P}=1.5\times 10^5$.}  
%	\label{fig:rg5}
%\end{figure}

%\begin{figure}
%	\includegraphics[width=0.8\textwidth]{difference}
%	\caption{Time series for the difference ($\delta O(t)-G\delta I(t)$) between scaled input $G\delta I(t)$ and the output $\delta O(t)$ from KMC simulations of the enzymatic push-pull loop, for $\lambda=100 (S=2.8\times 10^5 )$ and $\bar{P}=1.5\times 10^5$.}  
%%\end{figure}
%\begin{figure}
%	\includegraphics[width=0.9\textwidth]{nn.pdf}
%	\caption{Relative error $E$ vs. Phosphatase population with $\Lambda=100$ (Blue line, CL Theory: %Circles, Simulations and red line, WK optimal prediction).  }
%	\label{fig:error}
%\end{figure}

\end{document}